\documentclass[twocolumn]{aastex61}

\newcommand\aastex{AAS\TeX}

\shorttitle{\aastex\ Fast-pairwise neutrino oscillations associated with asymmetric $\nu$-emissions in CCSN}
\shortauthors{Nagakura et al.}

\begin{document}
\title{Fast-pairwise collective neutrino oscillations associated with asymmetric neutrino emissions in core-collapse supernova}
\correspondingauthor{Hiroki Nagakura}
\email{hirokin@astro.princeton.edu}

\author{Hiroki Nagakura}
\affiliation{Department of Astrophysical Sciences, Princeton University, Princeton, NJ 08544, USA}

\author{Taiki Morinaga}
\affiliation{Department of Science and Engineering, Waseda University, 3-4-1 Okubo, Shinjuku, Tokyo 169-8555, Japan}

\author{Chinami Kato}
\affiliation{Department of Aerospace Engineering, Tohoku University, 6-6-01 Aramaki-Aza-Aoba, Aoba-ku, Sendai 980-8579, Japan}

\author{Shoichi Yamada}
\affiliation{Department of Science and Engineering, Waseda University, 3-4-1 Okubo, Shinjuku, Tokyo 169-8555, Japan}
\affiliation{Advanced Research Institute for Science \&
Engineering, Waseda University, 3-4-1 Okubo,
Shinjuku, Tokyo 169-8555, Japan}

\begin{abstract}
We present a linear stability analysis of the fast-pairwise neutrino flavor conversion based on a result of our latest axisymmetric core-collapse supernova (CCSN) simulation with full Boltzmann neutrino transport. In the CCSN simulation, coherent asymmetric neutrino emissions of electron-type neutrinos ($\nu_{\rm e}$) and their anti-particles ($\bar{\nu}_{\rm e}$), in which the asymmetry of $\nu_{\rm e}$ and $\bar{\nu}_{\rm e}$ is anti-correlated with each other, occur at almost the same time as the onset of aspherical shock expansion. We find that the asymmetric neutrino emissions play a crucial role on occurrences of fast flavor conversions. The linear analysis shows that unstable modes appear in both pre- and post-shock flows; for the latter they appear only in the hemisphere of higher $\bar{\nu}_{\rm e}$ emissions (the same hemisphere with stronger shock expansion). We analyze in depth the characteristics of electron-lepton-number (ELN) crossing by closely inspecting the angular distributions of neutrinos in momentum space. The ELN crossing happens in various ways, and the property depends on the radius: in the vicinity of neutron star, $\bar{\nu}_{\rm e}$ ($\nu_{\rm e}$) dominates over $\nu_{\rm e}$ ($\bar{\nu}_{\rm e}$) in the forward (backward) direction: at the larger radius the ELN crossing occurs in the opposite way. We also find that the non-radial ELN crossing occurs at the boundary between no ELN crossing and the radial one, which is an effect of genuine multi-D transport. Our findings indicate that the collective neutrino oscillation may occur more commonly in CCSNe and suggest that the CCSN community needs to accommodate these oscillations self-consistently in the modelling of CCSNe.
\end{abstract}








\section{Introduction} \label{sec:intro}
More than three decades have passed since neutrinos emitted from SN1987A, a core-collapse supernova (CCSN) in the Large Magellanic Cloud, were directly detected by Kamiokande \citep{1987PhRvL..58.1490H} and IMB \citep{1987PhRvL..58.1494B}. Those neutrinos were produced deep inside the stellar core during the development of explosion. A proto-neutron star (PNS) is supposed to be formed and most of its internal energy ($\sim 10^{53} {\rm erg}$) was radiated by $10-30 {\rm MeV}$ neutrinos. This agrees qualitatively with the CCSN theory although the sparse data sample were insufficient to unveil the explosion mechanism.

Significant progresses have been made in the observational instruments, which will enable us to detect neutrinos from CCSNe with much higher statistics or at longer distances than those of SN1987A (see, e.g., \citet{2019arXiv190409996S} and reference therein). Super-Kamiokande, one of the operating neutrino detectors, is capable of detecting $\sim 10^4$ neutrinos for a Galactic CCSN and a few neutrinos for an event in Andromeda (M31) at the distance of $\sim 770$kpc (see, e.g., \citet{2011NuPhS.221..218R}). Hyper-Kamiokande, one of the next generation detector, will improve the sensitivity with an order of magnitude \citep{2011arXiv1109.3262A,2018arXiv180504163H}. The neutrino detections from multiple CCSNe or from a single CCSN event but with very high statistics will provide us vital information for comprehensive understanding of the CCSN mechanism.

On the theoretical side, tremendous progress has been also made very recently, for instance, three-dimensional CCSN simulations with spectral neutrino transport \citep{2012ApJ...749...98T,2015ApJ...807L..31L,2015ApJ...808L..42M,2014ApJ...786...83T,2016MNRAS.461L.112T,2016ApJS..222...20K,2016ApJ...831...98R,2017MNRAS.472..491M,2018ApJ...865...81O,2018ApJ...855L...3O,2018arXiv181105483M,2018ApJ...852...28S,2019MNRAS.482..351V,2019MNRAS.tmp..538B,2019arXiv190401699M,2019arXiv190408088N,2019arXiv190503786N} and those in axisymmetry but with multi-angle neutrino transport \citep{2008ApJ...685.1069O,2011ApJ...728....8B,2018ApJ...854..136N,2019ApJ...872..181H,2019arXiv190704863N} are nowadays available. Although the microphysics inputs, including neutrino-matter interactions and nuclear equation-of-state (EOS), were implemented in these simulations at different levels of refinement and accuracy, some of these simulations successfully reproduced explosions without artifices. In the upcoming exa-scale era, number of 3D CCSN simulations with approximate neutrino transport will be significantly increased, and 3D CCSN simulations with general relativistic full Boltzmann neutrino transport with further improved input physics will also become available \citep{2012PTEP.2012aA301K}. Both approaches will be complementary to each other in our efforts to make CCSN modeling more realistic.


Given the neutrino emissions from the core either by numerical simulations or by some simplified models, one may be able to calculate the expected signals on terrestrial detectors by taking into account neutrino oscillations (see e.g., \citet{2000PhRvD..62c3007D}). Neutrinos should experience the ordinary vacuum oscillation in the intervening space and will also go through the so-called Mikheyev-Smirnov-Wofenstein (MSW) resonance in the stellar envelope. Since neutrino signals from CCSN depend sensitively on the neutrino mass hierarchy (but they may be insensitive to the neutrino oscillation itself at later times in the neutron-star (NS) cooling \citep{2019arXiv190409996S}), the future detections of supernova neutrinos may reveal the ordering of neutrino masses. As such, the connection between theory and observation will be more tight in the next decades towards the comprehensive understanding of neutrino physics (see, e.g., \citet{2016MNRAS.461.3296N,2018MNRAS.480.4710S} for more details). 

Unfortunately, however, there remains a crucial concern in establishing realistic templates of neutrino signals and theoretical modelings of CCSN, that is, collective neutrino oscillations or the oscillations induced by neutrinos themselves. Even the most up-to-date simulations neglect these effects despite they may have an impact on both the explosion mechanism and the neutrino signals. There are mainly two reasons for the defect: (1) it is still uncertain whether the collective neutrino oscillations really occur in CCSNe or not; (2) if they do indeed, the treatment in CCSN simulations is not easy because of the disparity in scales and the nonlinearity of the phenomenon. Nevertheless, considerable efforts have been made in the CCSN community to address these issues by using various approaches (see recent reviews, e.g., \citet{2016NuPhB.908..366C,2016NCimR..39....1M,2018JPhG...45d3002H}, and references therein). The first issue has been studied by the linear stability analysis (see e.g., \citet{2017PhRvL.118b1101I,2018PhRvD..98j3001D,2018arXiv181206883A,2019PhRvD..99j3011D,2019PhRvD..99f3005Y}) or searching for the so-called electron-lepton-number (ELN) crossing \citep{2017ApJ...839..132T}, in which the energy-integrated angular distributions of $\nu_{\rm e}$ and $\bar{\nu}_{\rm e}$ in momentum space intersect with each other. Note that the ELN crossing is supposed to be a necessary condition for occurrences of the fast flavor conversion, one of the collective oscillations modes. The second issue has been, on the other hand, addressed by solving non-linear quantum kinetic equations under many simplifications (see e.g., \citet{2019arXiv190300022R,2019PhLB..790..545A}).

In this paper we tackle the former issue, focusing on fast flavor conversions \citep{2005PhRvD..72d5003S,2016PhRvL.116h1101S,2016JCAP...03..042C,2017PhRvL.118b1101I,2017PhRvD..96d3016C,2017JCAP...02..019D,2018PhRvD..98d3014A,2018PhRvD..97b3017D,2018PhRvD..98j3001D,2018JCAP...12..019A,2019PhLB..790..545A,2019PhRvD..99j3011D}. It should be noted that our CCSN simulations are capable of assessing occurrences of fast flavor conversion, since it feeds on the difference in the angular distributions among different species of neutrinos in momentum space, which is accessible only to the multi-angle neutrino transport like ours \citep{2018ApJ...854..136N,2019ApJ...872..181H,2019arXiv190704863N}. The methodology in this study is essentially the same as that in \citet{2019PhRvD..99j3011D}. We carry out the linear stability analysis as post-processing for the results of CCSN simulation but in this paper we employ one of the latest CCSN models, in which stronger asymmetric neutrino emissions ($\sim 10\%$) were observed to be associated with PNS kick. More interestingly, the asymmetries of $\nu_{\rm e}$ and $\bar{\nu}_{\rm e}$ emissions are anti-correlated with each other in this model, that is, the higher $\nu_{\rm e}$ emissions occur the opposite direction to the higher $\bar{\nu}_{\rm e}$ emissions\footnote{This characteristics is similar as that in LESA (lepton-emission self-sustained asymmetry) but the driving mechanism is different. See \citet{2019arXiv190704863N} for more details.}. We expect that such an anti-correlation will give an impact on fast flavor conversions and, indeed, find its positive sign unlike in the previous paper \citep{2019PhRvD..99j3011D}.


This paper is organized as follows. In Sec~\ref{sec:sumCCSNsimu} we briefly summarize our numerical modeling of CCSN. The basic equations of the linear stability analysis are given in Sec.~\ref{sec:linanaEq}, and then we present our main results in Sec.~\ref{sec:results}. Finally we conclude the paper with a summary in Sec.~\ref{sec:summary}. Unless otherwise stated, we use the unit with $c = G = \hbar = 1$, in which $c$, $G$, and $\hbar$ are the light speed, the gravitational constant, and the reduced Planck constant, respectively. We use the metric signature of $- + + +$. Greek and Latin indices run over $0–3$ and $1–3$, respectively.

\section{CCSN model} \label{sec:sumCCSNsimu}

\begin{figure*}
  \begin{minipage}{1.0\hsize}
        \includegraphics[width=\columnwidth]{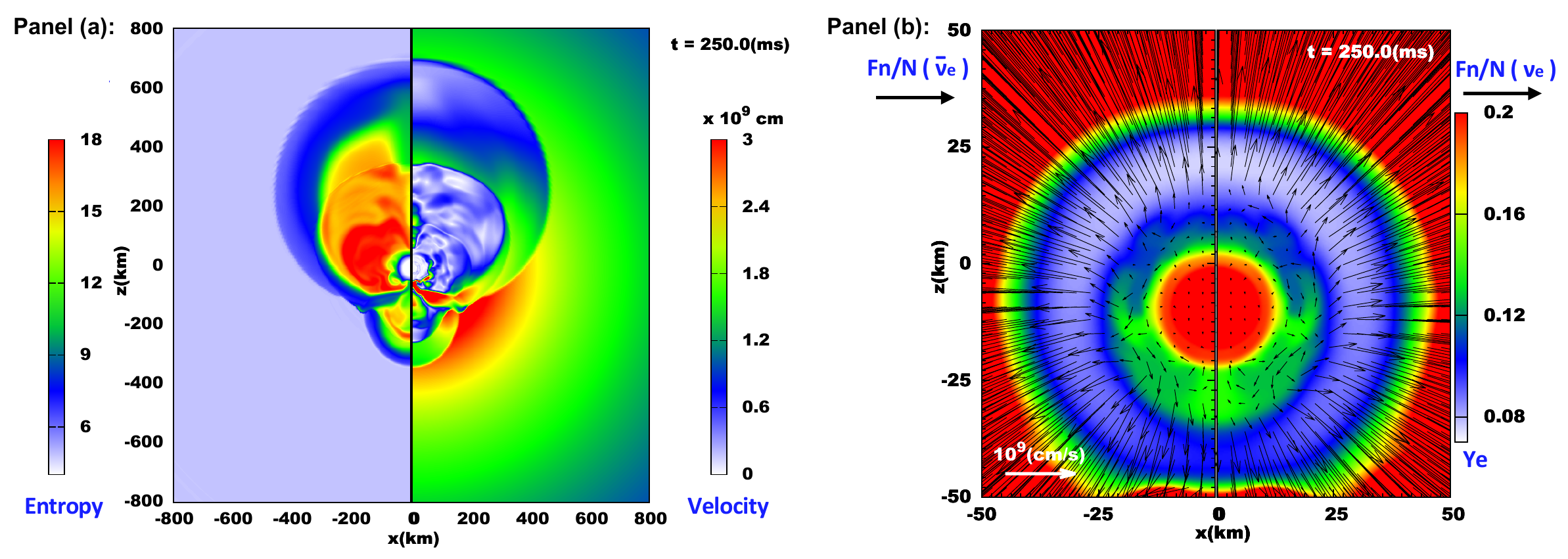}
    \caption{Panel (a): Color contours of entropy per baryon (left) and fluid speed (right) at $T_{\rm b} = 250$ms. Panel (b): Color contours of entropy per baryon with vector fields of $\bar{\nu}_{\rm e}$ (left) and $\nu_{\rm e}$ (right) number flux normalized by the each number density (right), respectively. Note that the spatial scale in each panel is different.}
    \label{2DHydrosnap}
  \end{minipage}
\end{figure*}

Neutrino distribution functions, $f_{\nu}$, as solutions of classical Boltzmann equations are fundamental quantities for the linear stability analysis of the fast flavor conversions (see Sec.~\ref{sec:linanaEq} for more details). In our CCSN simulations, we solve the Boltzmann equations for neutrino transport self-consistently but neglecting possible neutrino oscillations entirely. We use the neutrino data as the background (fixed point) for the linear stability analysis (see Sec.~\ref{sec:linanaEq} for more details). In this section we give an overview of our latest CCSN simulation.


 The details of the code development of our Boltzmann solver are described in a series of papers \citep{2012ApJS..199...17S,2014ApJS..214...16N,2017ApJS..229...42N,2019arXiv190610143N} and its reliability has been well established by a detailed comparison to another Monte-Carlo neutrino transport code \citep{2017ApJ...847..133R}. Some results of axisymmetric CCSN simulations by using our code can be seen in \citet{2018ApJ...854..136N,2019ApJ...872..181H,2019arXiv190704863N}.

In the present study, we employ the result of one of our latest axisymmetric CCSN models in \citet{2019arXiv190704863N}. In the simulation, the initial condition of the matter profile is taken from a 11.2 $M_{\sun}$ progenitor model in \citet{2002RvMP...74.1015W}, and the most up-to-date version of our code was run. The Boltzmann solver for neutrino transport is the same as that used in \citet{2018ApJ...854..136N}, while we recently improved input physics \citep{2019ApJS..240...38N} under a multi-nuclear variational method (VM) EOS \citep{2017JPhG...44i4001F}. The homogeneous nuclear matter is treated with the variational method \citep{2013NuPhA.902...53T,2017NuPhA.961...78T}, in which Argonne v18 \citep{1995PhRvC..51...38W} and UIX \citep{1983NuPhA.401...59C,1995PhRvL..74.4396P} potentials are adopted for the two- and three-body potentials, respectively. Inhomogeneous matter composed of various nuclei and dripped nucleons in nuclear statistical equilibrium are handled with various finite-density and thermal effects (see \citet{2017PhRvC..95b5809F} for more details). Based on the nuclear abundances provided by this EOS, we constructed new weak interaction tables that includes electron captures by heavy and light nuclei and positron captures by light nuclei. 

Below, we briefly summarize some characteristics in this CCSN model that deserve some mentions. As shown in Fig.~\ref{2DHydrosnap}, the shock wave expands strongly in the northern hemisphere (see Panel~(a)) and the PNS receives a linear momentum in the opposite direction (see Panel~(b)). Note that our code is capable of treating the PNS proper motion directly and self-consistently in real time. We also find that strong asymmetric emissions of electron-type neutrino ($\nu_{\rm e}$) and its anti-particle ($\bar{\nu}_{\rm e}$) occur at almost the same time as the onset of the aspherical shock expansion, which seems to be associated with the PNS kick \citep{2019arXiv190704863N}. The emissions of $\bar{\nu}_{\rm e}$ are higher in the hemisphere of stronger shock expansion (i.e., the northern hemisphere) whereas the $\nu_{\rm e}$ emissions have the opposite trend, i.e., higher in the hemisphere, into which the PNS is kicked (see also the upper panel of Fig.3 in \citet{2019arXiv190704863N}).

The asymmetric $\nu_{\rm e}$ and $\bar{\nu}_{\rm e}$ emissions are mainly caused by the non-spherical distributions of $Y_e$ around $10 \lesssim r \lesssim 25$km (see Panel~(b)), which are sustained by coherent lateral motions of matter. Interestingly, the linear momentum carried by neutrinos including the heavy-leptonic neutrino ($\nu_{\rm x}$) contributions is comparable to that of the PNS proper motion up to $300$ms after the bounce, which indicates that the asymmetric neutrino emissions play an important role in the acceleration of the PNS. We refer the reader to \citet{2019arXiv190704863N} for more details. As shown below, these asymmetric neutrino emissions are also important for the fast flavor conversion; indeed, higher $\nu_{\rm e}$ ($\bar{\nu}_{\rm e}$) emissions stabilize (trigger) the fast flavor conversion in the post-shock flows, the detail of which will be discussed in Sec.~\ref{sec:results}.

\section{Linear stability analysis of fast flavor conversion} \label{sec:linanaEq}
We conduct a linear stability analysis by employing the dispersion relation (DR) approach \citep{2017PhRvL.118b1101I}, which is probably one of the most convenient methods for the stability analysis in the literature. Below, we derive the DR of the fast flavor conversion. We refer readers to \citet{2017PhRvL.118b1101I,2017PhRvD..96d3016C,2018arXiv181206883A,2018JCAP...12..019A,2019PhRvD..99f3005Y} for more details.

We start with the equation of motion (EOM) for neutrino,
\begin{eqnarray}
&& i v^{\mu} \partial_{\mu} \rho_{\nu} = [H, \rho_{\nu}], \label{eq:basiceq}
\end{eqnarray}
where $v^{\mu}$, $\rho_{\nu}$ and $H$ denote the neutrino four velocity ($v^{\mu} = (1, \mbox{\boldmath $v$})$), the density matrix of neutrinos and the Hamiltonian matrix, respectively. Equation~(\ref{eq:basiceq}) is expressed on the flavor-basis, i.e., the diagonal components of the density matrix correspond to the distribution functions of flavor eigenstates, whereas the flavor coherence is expressed in the form of off-diagonal elements. The EOM of the anti-neutrinos can be included in Eq.~(\ref{eq:basiceq}) by using the flavor isospin convention, in which the density matrix of anti-neutrinos has an opposite sign of that of neutrinos and corresponds to negative frequencies (energies).


It should be noted that several simplifications have been done in Eq.~(\ref{eq:basiceq}): we take the ultra-relativistic limit; the spacetime is flat; we ignore the collision term of classical Boltzmann equations and the spin coherence \citep{2017PhRvD..95f3004T}. We further impose two-flavor approximation, in which we consider two flavors alone: $\nu_{\rm e}$ and $\nu_x$. Note that our numerical setup of CCSN simulations is compatible with the two-flavor approximation, in which we assumed that the neutrino distribution functions were identical among all heavy-leptonic neutrinos ($\nu_{\mu}$, $\nu_{\tau}$). Further studies are required to assess the impact of these simplifications, but are beyond the scope of this paper. Notwithstanding these uncertainties, Eq.~(\ref{eq:basiceq}) seems to contain the primary terms for the fast flavor conversion in CCSNe.
%


We decompose the Hamiltonian in Eq.~(\ref{eq:basiceq}) into three contributions,
\begin{eqnarray}
&& H = H_{V} + H_{M} + H_{\nu}, \label{eq:TotHamiltonian}
\end{eqnarray}
where
\begin{eqnarray}
&& H_{V} \equiv \frac{M^2}{2E} , \nonumber\\
&& H_{M} \equiv - v^{\mu} \Lambda_{\mu} \frac{\sigma_3}{2} , \nonumber \\
&& H_{\nu} \equiv - \sqrt{2} G_{F} \int \frac{E^{\prime 2} dE^{\prime}}{2 \pi^2} d\Gamma^{\prime} v^{\mu} v^{\prime}_{\mu} \rho^{\prime}_{\nu} . \label{eq:defHamiltonianComp}
\end{eqnarray}
From top to bottom they represent the vacuum, matter and neutrino-self-interaction contributions, respectively; $M^2$, $\sigma_3$, $G_F$ denote the mass-squared matrix, the third Pauli matrix and the Fermi constant, respectively, $\Lambda_{\mu}$ represents the matter potential in a covariant form, which can be written as
\begin{eqnarray}
&& \Lambda_{\mu} \equiv \sqrt{2} G_{F} (n_{e^{-}} - n_{e^{+}}) u_{\mu} , \label{eq:Sigmadef} 
\end{eqnarray}
where $n_{e^{-}}$, $n_{e^{+}}$ and $u^{\mu}$ are the electron- and positron number densities and their four velocity, respectively. Note that we have already subtracted the trace part of the matter potential, which does not affect the flavor conversion. In Eq.~(\ref{eq:Sigmadef}) muon and tau contributions are neglected, which may be good approximations in supernova core (but see \citet{2017PhRvL.119x2702B}). Following the common practice, we divide the integral in momentum space\footnote{The integral domain for the neutrino energy is from negative to positive infinity in Eq.~(\ref{eq:defHamiltonianComp}), since we take the flavor isospin convection.} into the energy part ($E^{2} dE/(2 \pi^2)$) and the angular one ($d\Gamma$) in the expression of $H_{\nu}$, where $E$ denotes the neutrino energy and $d\Gamma$ corresponds to the measure for the solid angle normalized by $4 \pi$ ($d\Gamma = d\mbox{\boldmath $v$}/4 \pi$). Hereafter, we ignore the vacuum contribution ($H_V$) since we focus only on the fast mode in the neutrino flavor conversions\footnote{Note that the vacuum contribution may play an important role as a seed perturbation to trigger the flavor conversion. We also refer the reader to \citet{2018JCAP...12..019A} for the case where slow- and fast-modes mix.}. Then Eq.~(\ref{eq:basiceq}) becomes energy-independent and one can integrate out the energy dependence. The energy-integrated form of the EOM can be written as
\begin{eqnarray}
&& i v^{\mu} \partial_{\mu} \mbox{\boldmath $\rho$}_{\nu} = [H, \mbox{\boldmath $\rho$}_{\nu}], \label{eq:basiceqEneint}
\end{eqnarray}
where
\begin{eqnarray}
&&  \mbox{\boldmath $\rho$}_{\nu} \equiv \frac{1}{2 \pi^2} \int_{-\infty}^{\infty} \rho_{\nu} E^2 dE, \label{eq:rhoeneinteg}
\end{eqnarray}
and $H_{\nu}$ defined in Eq.~(\ref{eq:defHamiltonianComp}) can be also rewritten in terms of $\mbox{\boldmath $\rho$}_{\nu}$ as
\begin{eqnarray}
&& H_{\nu} \equiv \sqrt{2} G_{F} \int d\Gamma^{\prime} v^{\mu} v^{\prime}_{\mu} \mbox{\boldmath $\rho^{\prime}$}_{\nu}. \label{eq:selfintHamiintegrhoexp}
\end{eqnarray}

It is well known that the matter potential, which dominates the vacuum contribution in supernova core, suppresses the neutrino flavor conversion as long as the neutrino contribution is neglected \citep{1979PhRvD..20.2634W}. It is hence reasonable to use the neutrino distribution functions obtained in our CCSN simulation, which neglects the neutrino oscillations, as unperturbed states in the linear stability analysis. They are indeed fixed points in Eq.~(\ref{eq:basiceqEneint}).

For latter convenience, we decompose the energy-integrated density matrix into the trace- and traceless part:
\begin{eqnarray}
&& \mbox{\boldmath $\rho$}_{\nu} = \frac{ \mbox{\boldmath $f$}_{\nu_{\rm e}} + \mbox{\boldmath $f$}_{\nu_{\rm x}}  }{2} I
+
\frac{ \mbox{\boldmath $f$}_{\nu_{\rm e}} - \mbox{\boldmath $f$}_{\nu_{\rm x}}  }{2}
\left(
\begin{array}{ccc}
s_{v}& \hspace{2mm} S_{v} \\
S^{*}_{v}& \hspace{2mm} -s_{v} \\
\end{array} 
\right)
. \label{eq:densitymatDecompose}
\end{eqnarray}
The coefficients, $\mbox{\boldmath $f$}_{\nu}$, are related with the unperturbed distribution function $f_{\nu_i}$ as
\begin{eqnarray}
&& \mbox{\boldmath $f$}_{\nu_i} = \frac{1}{2 \pi^2} \int_{0}^{\infty} (f_{\nu_i} - {f}_{\bar{\nu}_i}) E^2 dE \label{eq:CoeffDistrif}.
\end{eqnarray}
Since we assume $f_{\nu_x} = \bar{f}_{\nu_x}$ in our CCSN simulations, we set $\mbox{\boldmath $f$}_{\nu_x} = 0$ in this study. Hence the unperturbed density matrix is expressed as
\begin{eqnarray}
&& \mbox{\boldmath $\rho$}_{\nu(b)} = \frac{ \mbox{\boldmath $f$}_{\nu_{\rm e}} }{2} I
+
\frac{ \mbox{\boldmath $f$}_{\nu_{\rm e}}  }{2}
\left(
\begin{array}{ccc}
1& \hspace{2mm} 0 \\
0& \hspace{2mm} -1 \\
\end{array} 
\right)
. \label{eq:densitymatDecompose_b}
\end{eqnarray}

We linearize Eq.~(\ref{eq:basiceqEneint}) assuming that the off-diagonal component is small ($S_{v} \ll 1$), to obtain the following equation for $S_v$:
\begin{eqnarray}
 i ( \partial_t &+& \mbox{\boldmath $v$} \cdot \mbox{\boldmath $\nabla$}_{r} ) S_{v} \nonumber \\
&&=  - v^{\mu} (\Lambda_{\mu} + \Phi_{\mu} ) S_{v} + \int d\Gamma^{\prime} v^{\mu} v^{\prime}_{\mu} G_{v^{\prime}} S_{v^{\prime}} , \label{eq:Slinear}
\end{eqnarray}
with
\begin{eqnarray}
&& G_{v} \equiv \sqrt{2} G_F \mbox{\boldmath $f$}_{\nu_{\rm e}} (\mbox{\boldmath $v$}), \label{eq:Gvdef} \\
&& \Phi_{\mu} \equiv  \int d\Gamma G_{v} v_{\mu}. \label{eq:defELNcurrent}
\end{eqnarray}
Note that the diagonal component remains conserved in the linear order (see also \citet{2018JCAP...12..019A}).

To obtain solutions of Eq.~(\ref{eq:Slinear}), we take a plane-wave ansatz, which can be written in the form
\begin{eqnarray}
S_{v} = Q_{v} {\rm exp}[-i( \Omega t - \mbox{\boldmath $K$} \cdot \mbox{\boldmath $r$}  )]. \label{eq:S_planewave} 
\end{eqnarray}
Then the EOM can be rewritten as
\begin{eqnarray}
v^{\mu} k_{\mu} Q_{v} = - \int d\Gamma^{\prime} v^{\mu} v^{\prime}_{\mu} G_{v^{\prime}} Q_{v^{\prime}}, \label{eq:S_planewave} 
\end{eqnarray}
where $k_{\mu} (=(-\omega, \mbox{\boldmath $k$} )) \equiv K_{\mu} - \Lambda_{\mu} - \Phi_{\mu} $ with $K_{\mu} = (-\Omega, \mbox{\boldmath $K$} )$. We can further rewrite the equation as
\begin{eqnarray}
Q_{v} = \frac{v^{\mu} a_{\mu}}{ v^{\gamma} k_{\gamma} }, \label{eq:Qvsimple}
\end{eqnarray}
where $a_{\mu}$ is defined as
\begin{eqnarray}
a_{\mu} \equiv - \int d\Gamma v_{\mu} G_{v} Q_{v}, \label{eq:defpolarizationvec}
\end{eqnarray}
which is called the polarization vector. Inserting Eq.~(\ref{eq:Qvsimple}) into the right hand side of Eq.~(\ref{eq:defpolarizationvec}), we obtain the following relation
\begin{eqnarray}
\Pi^{\mu \nu} a_{\nu} = 0, \label{eq:lastLinearEq1}
\end{eqnarray}
where
\begin{eqnarray}
\Pi^{\mu \nu} &\equiv& \eta^{\mu \nu} + \int d\Gamma G_{v} \frac{ v^{\mu} v^{\nu} }{v^{\gamma} k_{\gamma}} \nonumber \\
&=&  \eta^{\mu \nu} - \int d\Gamma G_{v} \frac{ v^{\mu} v^{\nu} }{ \omega - \mbox{\boldmath $v$} \cdot \mbox{\boldmath $k$} }. \label{eq:lastLinearEq2}
\end{eqnarray}
In this equation, $\eta^{\mu \nu} = {\rm diag} (-1,1,1,1)$ is the Minkowski metric and $\Pi^{\mu \nu}$ is called the polarization tensor. The nontrivial solutions can be obtained only when
\begin{eqnarray}
{\rm det} \hspace{0.5mm} \Pi = 0, \label{eq:DispRelation}
\end{eqnarray}
which gives a relation between $\omega$ and $\mbox{\boldmath $k$}$ or the DR.


We numerically search the solutions of Eq.~(\ref{eq:DispRelation}) that give instability. As is well known, however, we need care in numerically finding these solutions, since the so-called spurious modes \citep{2012PhRvD..86l5020S} are artificially generated if we conduct integrations numerically by discritization. More recently, two of the authors of this paper developed a novel method to avoid this unpleasant issue, in which the integrations are done analytically with some basis functions \citep{2018PhRvD..97b3024M}. The validity of the method was confirmed in our previous paper \citep{2019PhRvD..99j3011D}. This method is a bit computationally costly, however, since high-order polynomials are required to compute accurately the DR for strongly forward-peaked angular distributions, and is not suitable for a survey of wide spatial regions in many snapshots. We hence use a simpler formula for the maximum growth rate of unstable solution (see also Eq.(8) in \citet{2019arXiv190913131M}):
\begin{eqnarray}
\max_{k\in\mathbb{R}^3} && \left( {\rm Im}(\omega(k)) \right) \nonumber \\
&\sim& \sqrt{ \left| \left(\int_{G_{v>0}} d\Gamma G_{v}\right)\left(\int_{G_{v<0}} d\Gamma G_{v}\right)   \right| }. \label{eq:approxiGrowth}
\end{eqnarray}
This approximate expression is partially motivated by the fact that unstable solutions appear when the ELN crossing occurs. Its validity will be checked at some selected points (see Sec.~\ref{subsec:stabi} for more details).

\begin{figure*}
  \rotatebox{90}{
  \begin{minipage}{1.3\hsize}
        \includegraphics[width=\columnwidth]{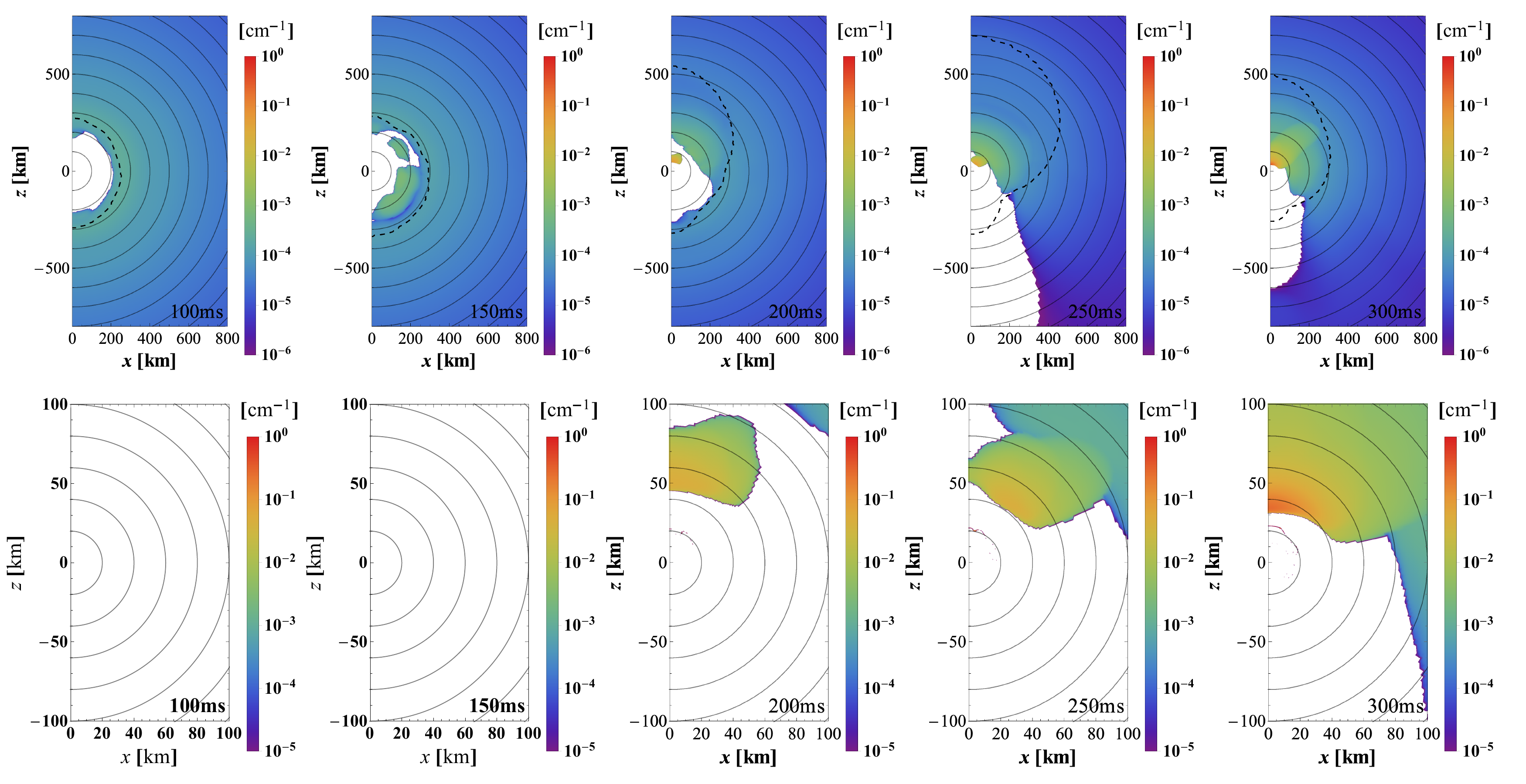}
    \caption{Color contours of the growth rate of the fast flavor conversion. From left to right panels, the times are $T_{\rm b} = 100, 150, 200, 250$ and $300$ms, respectively. The location of shock wave is also marked by a black dashed line.} The bottom panels focus on the region with $r \lesssim 100$km.
    \label{SpatialDistri_Growth}
  \end{minipage}}
\end{figure*}

\begin{figure*}
  \begin{minipage}{1.0\hsize}
        \includegraphics[width=\columnwidth]{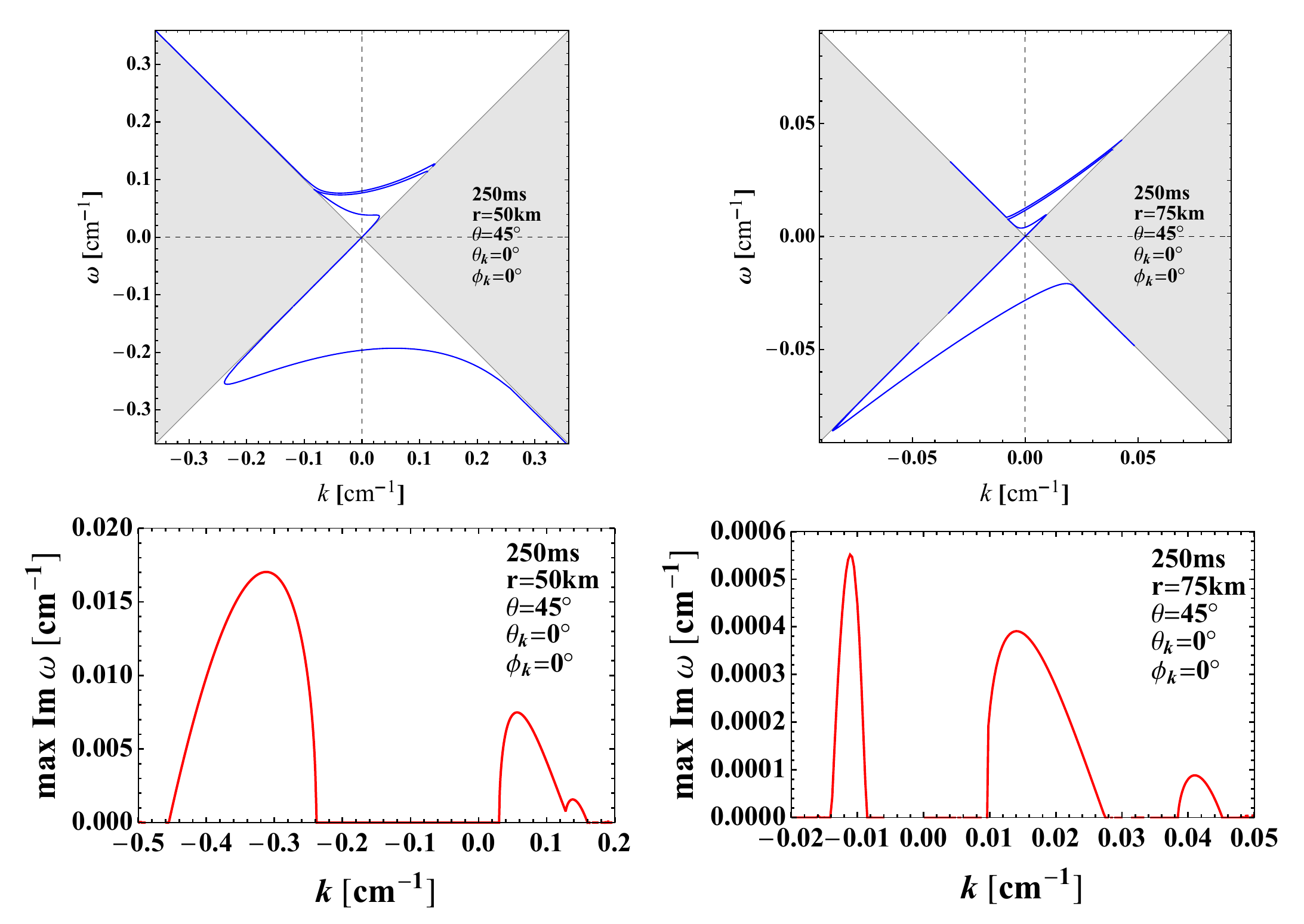}
    \caption{Top: Dispersion relations of fast flavor conversion with respect to the wave number of local radial direction at two representative unstable locations. The left panel displays the result at $r=50$km along a radial ray with $\theta=45^{\circ}$, while the right one is the same one as the left panel but for $r=75$km. The time is $T_{\rm b} = 250$ms. Bottom: The growth rate as a function of wave number. The spatial location and time in these panels are the same as those of the top panel in the same line. }
\label{DR_GRdetail}
  \end{minipage}
\end{figure*}

\section{Results} \label{sec:results}
In this section we present main results of this paper. We first summarize the stability analysis of the fast flavor conversion in Sec.~\ref{subsec:stabi}. Then we discuss the role of asymmetric neutrino emissions, paying particular attention to the ELN crossings in Sec~\ref{subsec:asymneutrino}.

\subsection{Stability of fast flavor conversion} \label{subsec:stabi}
Figure~\ref{SpatialDistri_Growth} displays spatial maps of the growth rate of the fast flavor conversion for some selected snapshots, in which colors other than white represent the unstable region. Note that the growth rate is computed based on Eq.~(\ref{eq:approxiGrowth}). One of the common properties among all these snapshots is that there exist unstable modes in the wide areas at large radii (see e.g., the pre-shock region with $r\gtrsim 200$km in the top panels of Fig.~\ref{SpatialDistri_Growth}). The occurrence of ELN crossing is also confirmed in the region: $\nu_{\rm e}$ is dominant over $\bar{\nu}_{\rm e}$ in the outgoing (${\rm cos} \hspace{0.5mm} \theta_{\nu} = 1$) direction whereas the trend is opposite in the backward (${\rm cos} \hspace{0.5mm} \theta_{\nu} = -1$) direction; this implies that the crossing occurs somewhere in between. Below, we describe the essence of the mechanism (but see \citet{2019arXiv190913131M} for more details).


Accreted matter in pre-shock region is mostly composed by heavy nuclei (see e.g., the right panel of Fig. 11 in \citet{2019ApJS..240...38N}). Some neutrinos emitted from PNS experiences scatterings by these nuclei and then turn their directions. Since the outgoing neutrinos are several orders of magnitude more abundant than those in the inward direction, the scattered neutrinos govern the neutrino population in the inward direction (see also Fig.2b in \citet{2019arXiv190913131M}). Note also that, since the average energy of $\bar{\nu}_{\rm e}$ is higher than that of $\nu_{\rm e}$, $\bar{\nu}_{\rm e}$ experiences more scatterings with nuclei than $\nu_{\rm e}$, which makes $\bar{\nu}_{\rm e}$ be more abundant than $\nu_{\rm e}$ in the inward direction. Thus, the ELN sign is negative in the inward direction, which is opposite to that in outgoing direction (see e.g., \citet{2017ApJ...839..132T}), i.e., the ELN crossing appears. 


It should be noted that the ELN crossing in the pre-shock region has been overlooked so far even in most recent papers (see, e.g., \citet{2018arXiv181206883A,2019PhRvD..99j3011D,2019arXiv190407236S}). There are probably some reasons for this. Almost every previous work has considered the possibility of fast flavor conversion only in the post-shock region, in particular, the vicinity of PNS \citep{2018arXiv181104215A,2019PhRvD..99j3011D}, where neutrinos are more abundant than in the pre-shock region. In addition to this, although the ELN property in the inward direction at pre-shock region is crucial ingredient for the fast flavor conversion, they have received little attention. This is because neutrinos have strongly forward-peaked angular distributions at large radii, which have eluded recognition of the ELN crossing. As pointed out in \citet{2019arXiv190913131M}, the ELN crossing is tiny but the growth rate is large enough to induce the flavor conversion, which may give an impact on terrestrial observations of CCSN neutrinos. Note that the neutrino signals in early post-bounce phase will not be affected by the fast flavor conversion, since $\bar{\nu}_{\rm e}$ emissions are much smaller than $\nu_{\rm e}$ at the phase. Indeed, we observe the ELN crossings from $T_{\rm b} \gtrsim 50$ms in this model.

In the post-shock regions, on the other hand, most heavy nuclei are broken up into lighter nuclei or nucleons; thus the above mechanism does not operate. In fact, $\nu_{\rm e}$ is dominant over $\bar{\nu}_{\rm e}$ for all flight directions up to $T_{\rm b} \sim 150$ms, and there is no positive sign of fast flavor conversion (see first and second panels from left on the bottom row in Fig.~\ref{SpatialDistri_Growth}.). This is qualitatively consistent with our previous paper \citep{2019PhRvD..99j3011D}. As shown in other plots on the bottom row in Fig.~\ref{SpatialDistri_Growth}, however, unstable modes appear in the northern hemisphere (the same hemisphere with stronger shock expansion) from $T_{\rm b} \gtrsim 200$ms, and persist throughout the late phase. The role of the asymmetric neutrino emissions in the fast flavor conversion will be discussed in detail in Sec.~\ref{subsec:asymneutrino}.




In the top row of Fig.~\ref{DR_GRdetail}, we show the DR for wave number vectors, which are chosen to be radial, at two unstable locations in the post-shock region. Equation~(\ref{eq:DispRelation}) is solved with spherical harmonics up to $\ell=9$. As we have already mentioned, the analytic integration with basis functions is the key to avoid spurious modes. In the figure, we find some peaks in the DR, which may be a good indicator for the existence of unstable models (See \citet{2019PhRvD..99j3011D} for more details). In the bottom panels, we show the growth rates as a function of the wave number. There exist unstable modes, indeed, as indicated by the DR as well as by the approximate prescription of Eq.~(\ref{eq:approxiGrowth}). 

\subsection{Role of asymmetric $\nu$ emissions} \label{subsec:asymneutrino}

\begin{figure}
  \begin{minipage}{1.0\hsize}
        \includegraphics[width=\columnwidth]{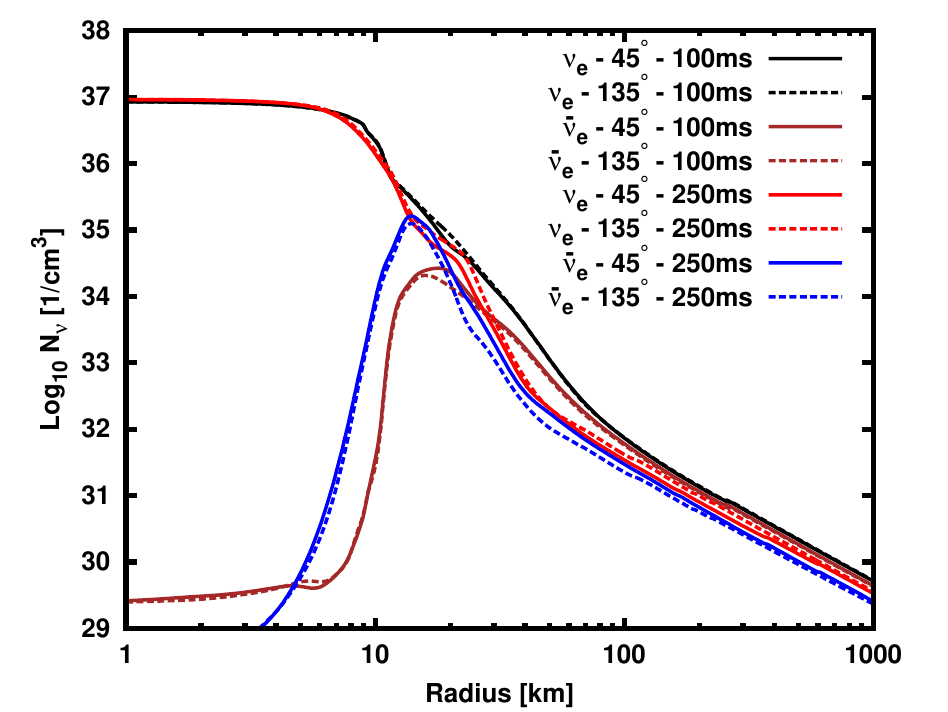}
    \caption{Distributions of neutrino number density ($N_{\nu}$) along two radial rays, $\theta = 45^{\circ}$ (solid lines) and $135^{\circ}$ (dashed lines), at two different snapshots, $T_{\rm b} = 100$ms and $T_{\rm b} = 250$ms.}
\label{graph_asymprofile}
  \end{minipage}
\end{figure}

\begin{figure}
  \begin{minipage}{1.0\hsize}
        \includegraphics[width=\columnwidth]{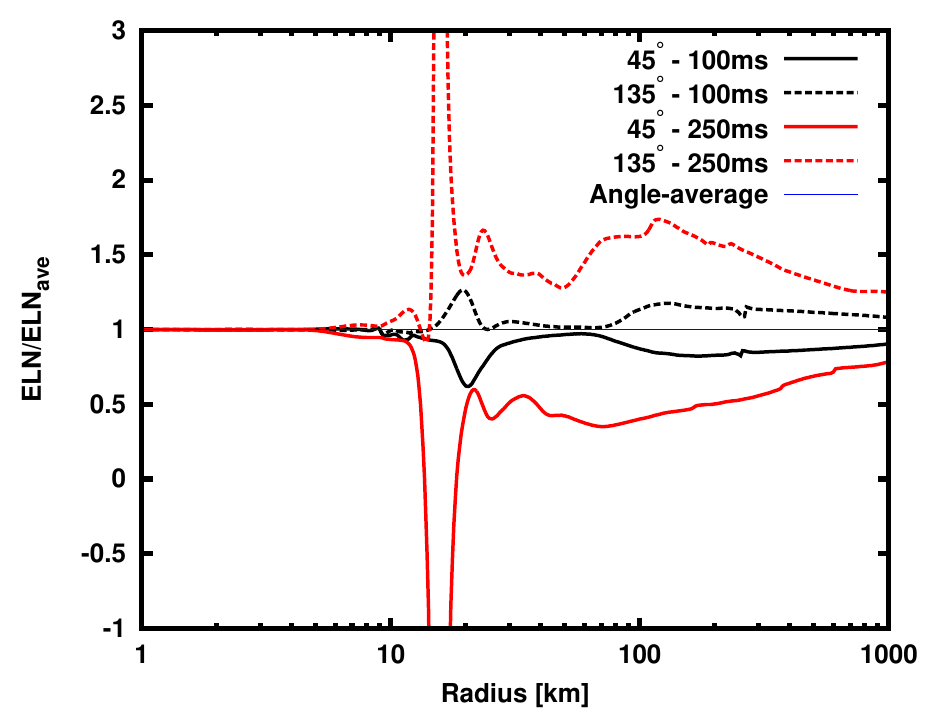}
    \caption{The radial profile of ELN asymmetry which is defined as the ratio of the ELN along each radial ray ($\theta = 45^{\circ}$ and $135^{\circ}$) to their spherical average. According to the defenition, the deviation from unity (the thin blue line in this figure) corresponds to the degree of asymmetries of ELN distributions.}
\label{graph_asymELNprofile}
  \end{minipage}
\end{figure}

\begin{figure}
  \begin{minipage}{1.0\hsize}
        \includegraphics[width=\columnwidth]{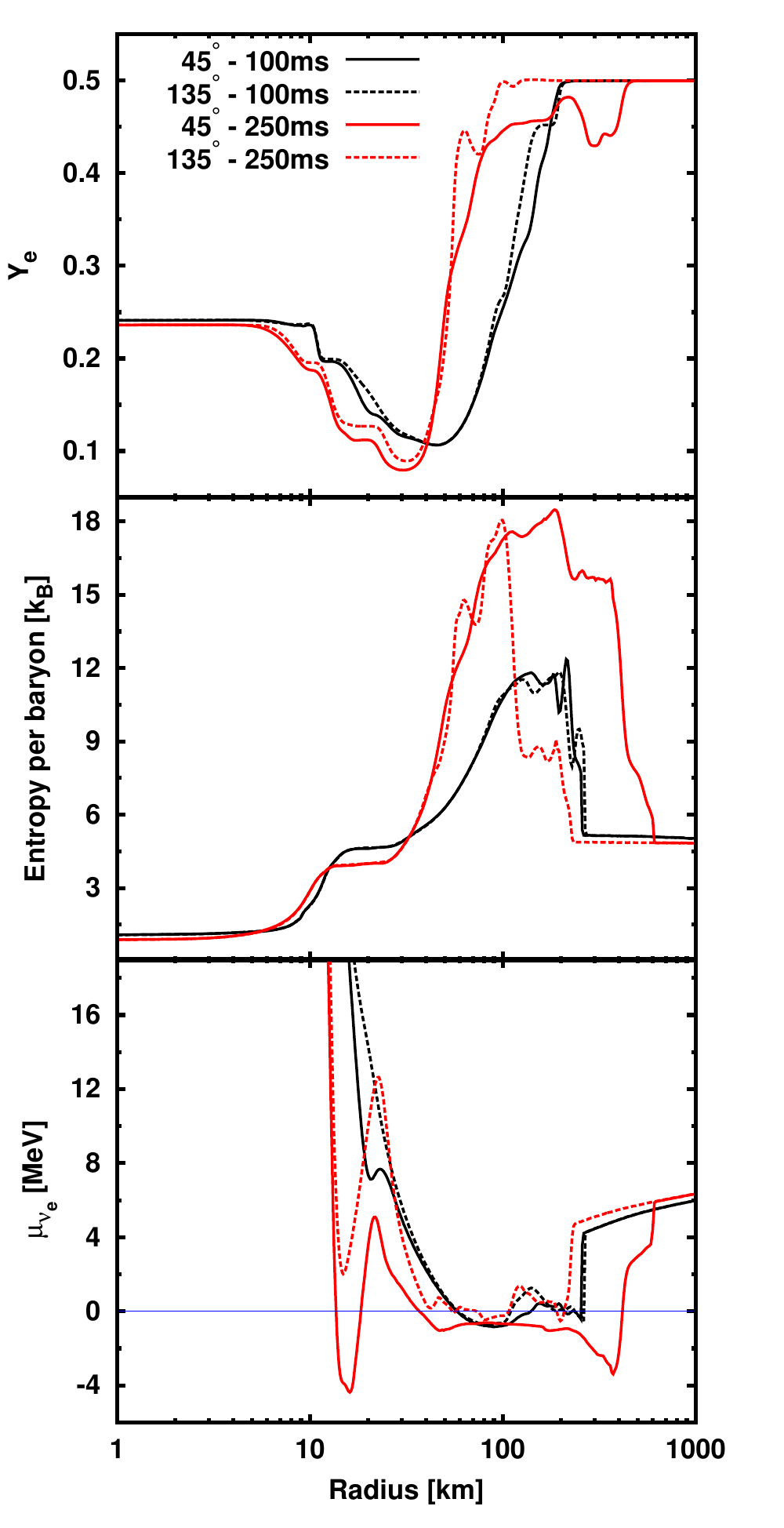}
    \caption{Radial profiles of electron fraction (top), entropy per baryon (middle) and chemical potential of electron-type neutrinos (bottom). The line type distinguishes the different radial ray with $\theta = 45^{\circ}$ (solid) and $\theta = 135^{\circ}$ (dashed), respectively. The black and red lines denote $T_{\rm b} = 100$ms and $T_{\rm b} = 250$ms.}
\label{graph_asymMatterprofile}
  \end{minipage}
\end{figure}

\begin{figure*}
  \begin{minipage}{1.0\hsize}
        \includegraphics[width=\columnwidth]{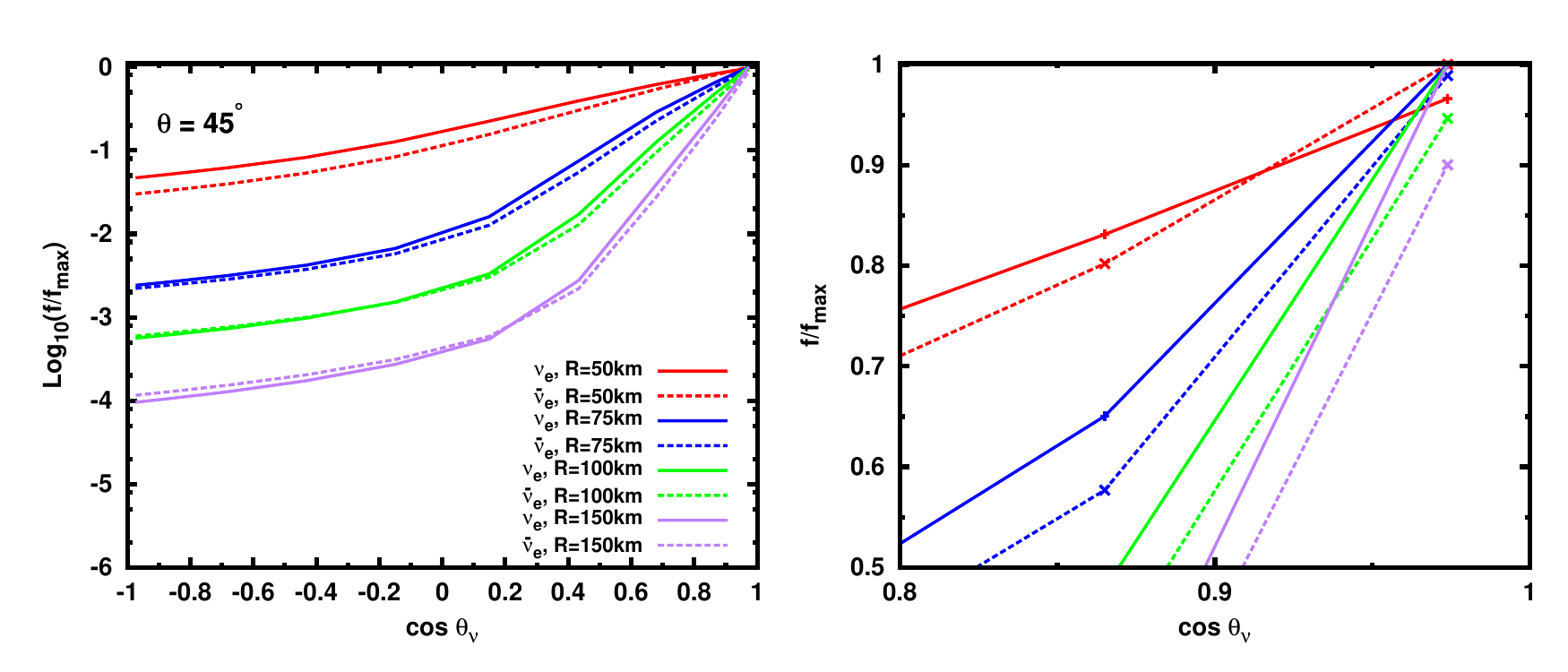}
    \caption{The left panel displays the $\phi_{\nu}$-integrated distribution functions of $\nu_{\rm e}$ (solid lines) and $\bar{\nu}_{\rm e}$ (dashed lines) as a function of $\theta_{\nu}$. The vertical axis is normalized by $f_{\rm max}$, which is defined as the maximum value of $f$ for both $\nu_{\rm e}$ and $\bar{\nu}_{\rm e}$ at the same spatial point. The right panel shows the same quantity but focuses on the forward direction (${\rm cos} \hspace{0.5mm} \theta_{\nu} > 0.8$). Different colors imply the different radial positions along the same radial ray with $\theta = 45^{\circ}$. The time is $T_{\rm b} = 250$ms.}
\label{graph_axisymmetric_crossingcheck_45deg_250ms}
  \end{minipage}
\end{figure*}

\begin{figure*}
  \begin{minipage}{1.0\hsize}
        \includegraphics[width=\columnwidth]{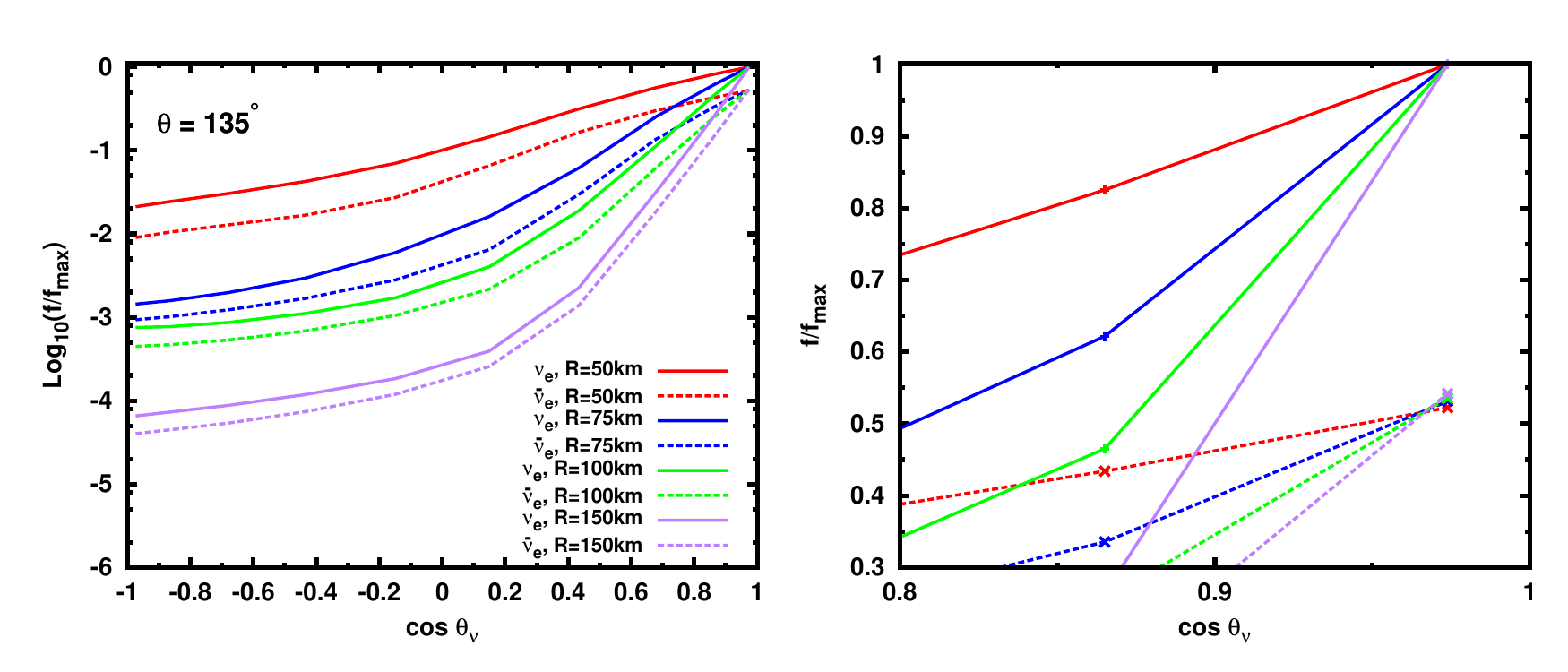}
    \caption{Same as Fig.~\ref{graph_axisymmetric_crossingcheck_45deg_250ms} but along a radial ray with $\theta = 135^{\circ}$.}
\label{graph_axisymmetric_crossingcheck_135deg_250ms}
  \end{minipage}
\end{figure*}

\begin{figure}
  \begin{minipage}{1.0\hsize}
        \includegraphics[width=\columnwidth]{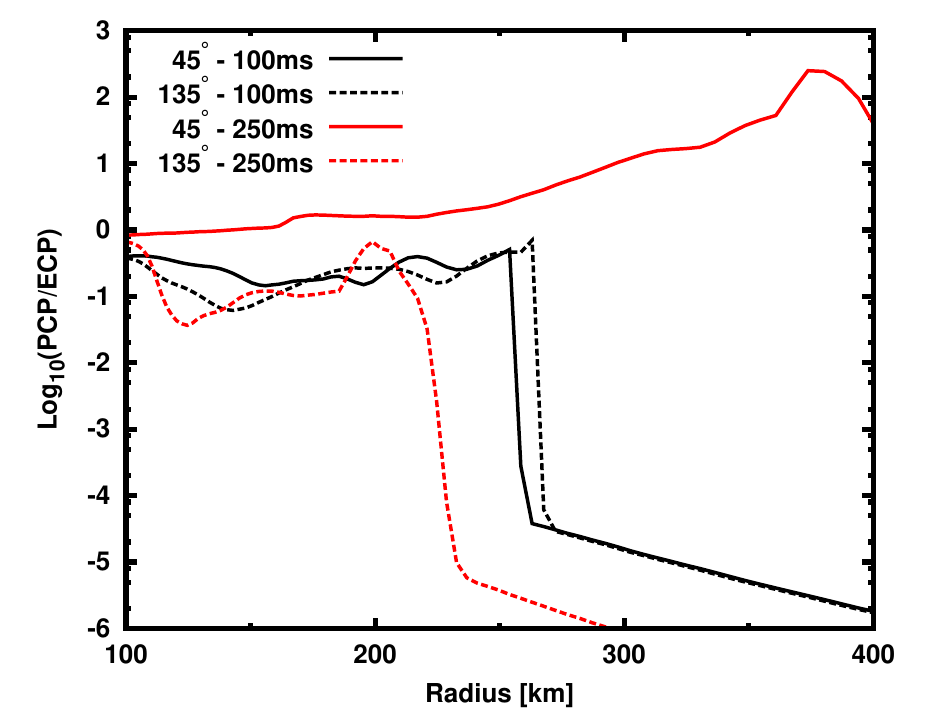}
    \caption{Radial profiles of the ratio of positron capture on free protons to electron captures on free neutrons. We display them along two different radial rays ($\theta = 45^{\circ}$ and $\theta = 135^{\circ}$) at two different snapshot ($T_{\rm b} = 100$ms and $T_{\rm b} = 250$ms). The line type and colors distinguish the radial ray and time, respectively, which are the same in Fig.~\ref{graph_asymMatterprofile}.}
\label{graph_CompareECP_PCP_Back}
  \end{minipage}
\end{figure}

\begin{figure*}
  \begin{minipage}{1.0\hsize}
        \includegraphics[width=\columnwidth]{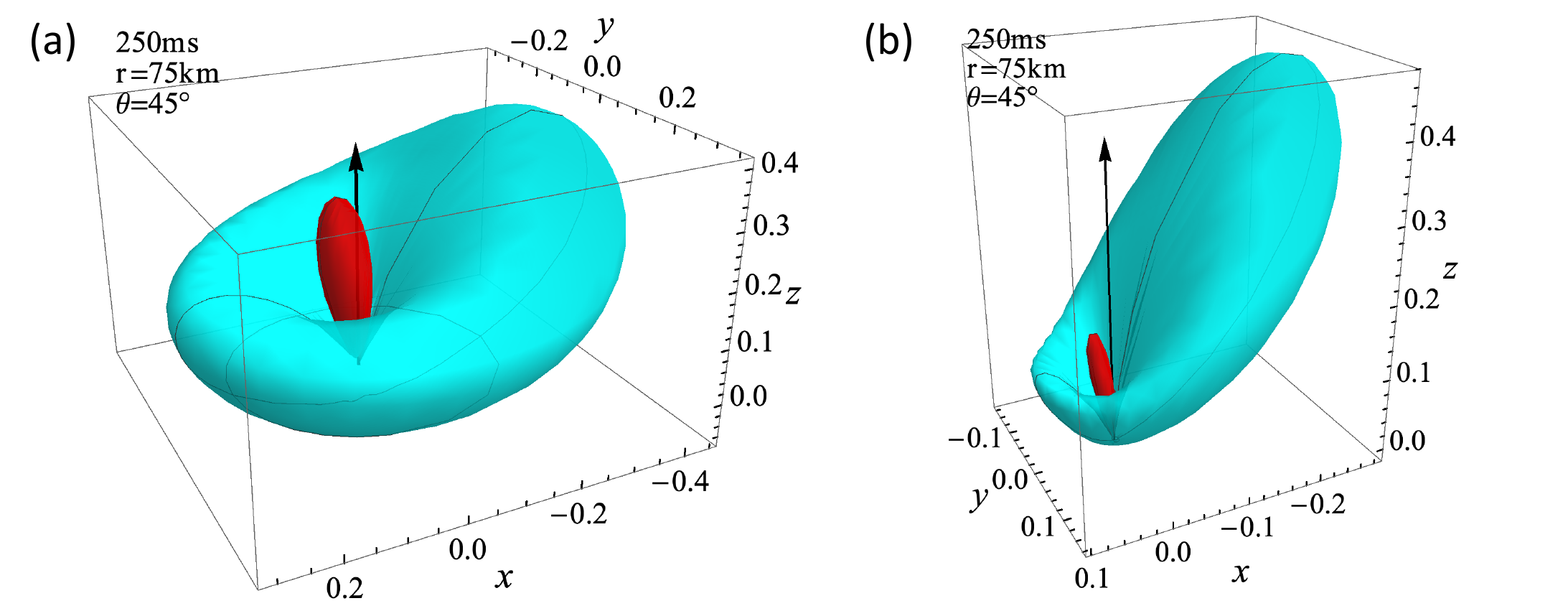}
    \caption{Angular distributions of electron-lepton-number (ELN) in momentum space. They are displayed for two different spatial locations; $r=50$km for the left and $r=75$km for the right panels, respectively, with the same radial ray ($\theta = 45^{\circ}$) and the same time snapshot ($T_{\rm b} = 250$ms). The cyan and red colors imply the positive and negative ELN, respectively. The arrow represents the local radial direction (along with z-direction in the panel) from the coordinate origin of momentum space.}
\label{3Dang}
  \end{minipage}
\end{figure*}

\begin{figure}
  \begin{minipage}{1.0\hsize}
        \includegraphics[width=\columnwidth]{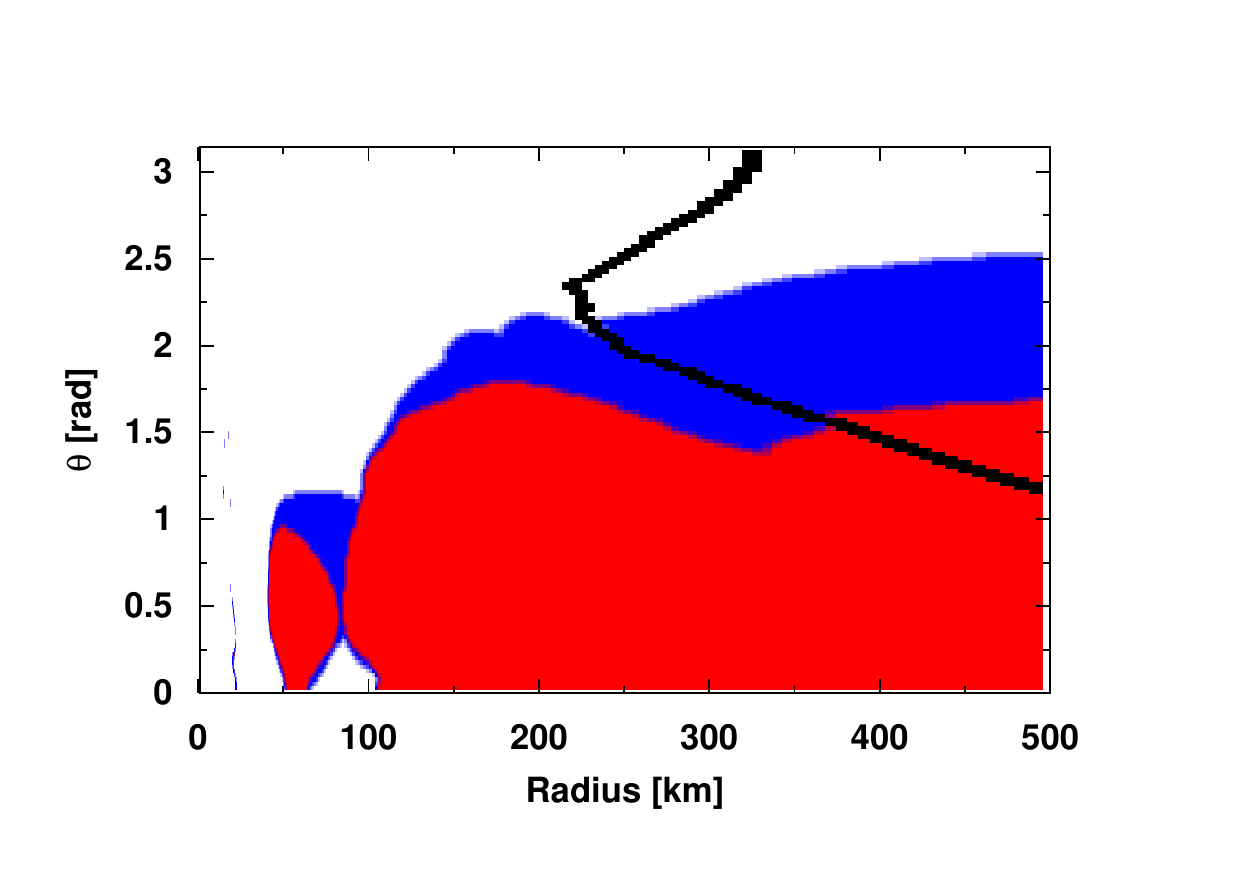}
    \caption{Color-coded 2D map ($r-\theta$ plane) to see the type of ELN crossing. Red and blue colors are radial and non-radial ELN crossing (see the main text for the definition of the radial and non-radial crossing), respectively. The shock radius is marked as a black solid line. The time is $T_{\rm b} = 250$ms.}
\label{graphCrosscheck}
  \end{minipage}
\end{figure}


Next we turn our attention to the role of the asymmetric neutrino emission in the occurrence of the fast flavor conversion. Figs.~\ref{graph_asymprofile} and \ref{graph_asymELNprofile} portray the asymmetry in neutrino emissions: the former displays the radial profile of number density of $\nu_{\rm e}$ ($N_{\nu_{\rm e}}$) and $\bar{\nu}_{\rm e}$ ($N_{\bar{\nu}_{\rm e}}$) along two selected radial rays ($\theta = 45$ and $135^{\circ}$) for two snapshots at $T_{\rm b} = 100$ and $250$ms: the latter displays their ELN asymmetry which is defined by the ratio of the ELN along each radial ray to the angle average. At $T_{\rm b} = 100$ms, the radial distributions of $N_{\nu_{\rm e}}$ and $N_{\bar{\nu}_{\rm e}}$ are roughly spherically symmetric except for the region of $15 \lesssim r \lesssim 30$km, in which violent matter motions produced by convections in PNS disturb the neutrino distributions. Occasionally $N_{\nu_{\rm e}}$ and $N_{\bar{\nu}_{\rm e}}$ become close each other (see, e.g., black and brown solid lines in Fig.~\ref{graph_asymprofile}). However, $N_{\nu_{\rm e}}$ is roughly one order of magnitude larger than $N_{\bar{\nu}_{\rm e}}$, and the ELN crossing hardly occurs\footnote{The angular distributions of $\nu_{\rm e}$ and $\bar{\nu}_{\rm e}$ are both nearly isotropic in this region.}. The dominance of $\nu_{\rm e}$ over $\bar{\nu}_{\rm e}$ can be understood through chemical potential of $\nu_{\rm e}$ ($\mu_{\nu_{\rm e}}$)\footnote{The chemical potential of $\nu_{\rm e}$ is defined as $\mu_{\nu_{\rm e}} \equiv \mu_{{\rm e}} + \mu_{{\rm p}} - \mu_{{\rm n}}$, where $\mu_{{\rm e}}$, $\mu_{{\rm p}}$ and $\mu_{{\rm n}}$ are that of electron, proton and neutron, respectively.}. At $T_{\rm b} = 100$ms, $\mu_{\nu_{\rm e}}$ is $\gtrsim 8$MeV in the region of $15 \lesssim r \lesssim 30$km (see black lines in the bottom panel of Fig.~\ref{graph_asymMatterprofile}), and $\nu_{\rm e}$ is more abundant than $\bar{\nu}_{\rm e}$. Although the difference of $N_{\nu_{\rm e}}$ and $N_{\bar{\nu}_{\rm e}}$ becomes smaller with increasing radius, $\nu_{\rm e}$ still dominates over $\bar{\nu}_{\rm e}$ in the above region. In such environments, the ELN crossing does not occur in the post-shock flows, which is qualitatively same results as those found in spherically symmetric CCSN simulations (see e.g., \citet{2017ApJ...839..132T}).



At $T_{\rm b} = 250$ms, the asymmetric neutrino emissions are noticeable: $\nu_{\rm e}$ at $\theta = 135^{\circ}$ is more abundant than at $\theta = 45^{\circ}$ (see red lines in Fig.~\ref{graph_asymprofile}), whereas $\bar{\nu}_{\rm e}$ has an opposite trend (see blue lines in Fig.~\ref{graph_asymprofile}). The characteristics of asymmetric neutrino emissions can be also seen in Fig.~\ref{graph_asymELNprofile}: the ELN distribution along the ray with $\theta = 45^{\circ} (135^{\circ})$ is $\gtrsim~50~\%$ lower (higher) than that of angle-average in the region between PNS and shock radii\footnote{The rapid spike of ELN asymmetry in the region between $10{\rm km} < r < 20{\rm km}$ is attributed to the fact that the angle-average ELN is almost zero.}. This indicates that the number densities of $\nu_{\rm e}$ and $\bar{\nu}_{\rm e}$ become close to each other along the radial ray with $\theta = 45^{\circ}$, whereas they are different markedly in the ray with $\theta = 135^{\circ}$.  As pointed out by \citet{2018arXiv181206883A}, the ELN crossings potentially occur if $\bar{\nu}_{\rm e}$-to-$\nu_{\rm e}$ ratio becomes close to unity, i.e., the radial ray with $\theta = 45^{\circ}$ is preferable for the ELN crossing.


The ELN crossing occurs at $r \gtrsim 50$km on this ray, indeed\footnote{We also find the ELN crossing at $r \sim 20$km occasionally (see, e.g., the bottom right panel of Fig.~\ref{SpatialDistri_Growth}). However, they may be due to the numerical noise; hence we do not discuss them in this paper.}. Fig.~\ref{graph_axisymmetric_crossingcheck_45deg_250ms} shows the angular distributions of $\nu_{\rm e}$ and $\bar{\nu}_{\rm e}$ along the radial ray with $\theta = 45^{\circ}$ but at different radii. In these plots, the $\phi_{\nu}$ dependence is integrated out. Note also that we normalize the vertical axis by $f_{\rm max}$ which is defined as the maximum value of $\phi_{\nu}$-integrated distribution functions for both $\nu_{\rm e}$ and $\bar{\nu}_{\rm e}$ at the same spatial point. As shown in these plots, there are ELN crossings at $r=50, 100$ and $150$km (but see below for the case with $r=75$km). At $r=50$km, $\bar{\nu}_{\rm e}$ is dominant over $\nu_{\rm e}$ for the forward direction (${\rm cos} \hspace{0.5mm} \theta_{\nu} > 0.9$), whereas the trend is opposite in other directions. For the other two radii ($r=100$ and $150$km), the ELN crossing occurs in the opposite way to that of $r=50$km. In Fig.~\ref{graph_axisymmetric_crossingcheck_135deg_250ms} we also display the same quantities but on a different radial ray ($\theta = 135^{\circ}$) for comparison. It clearly shows that $\nu_{\rm e}$ always dominates over $\bar{\nu}_{\rm e}$ and, indeed, the neutrino distributions are stable to the fast flavor conversion.


As mentioned above, how the ELN crossing occurs along the radial ray with $\theta = 45^{\circ}$ depends on the radius, which indicates that different mechanisms are responsible for the ELN crossings. At $r \sim 50$km, the angular distribution of $\bar{\nu}_{\rm e}$ is more forward-peaked than that of $\nu_{\rm e}$, since $\bar{\nu}_{\rm e}$ decouples from matter at a smaller radius than $\nu_{\rm e}$. Although $N_{\nu_{\rm e}}$ is larger than $N_{\bar{\nu}_{\rm e}}$ (see red and blue solid lines in Fig.~\ref{graph_asymprofile}), its difference is much smaller than that along the ray with $\theta = 135^{\circ}$. This is mainly due to the fact that $\nu_{\rm e}$ is more efficiently absorbed by neutrons and the positron capture is also facilitated in the lower $Y_e$ environment along the radial ray with $\theta = 45^{\circ}$ (see the top panel of Fig.~\ref{graph_asymMatterprofile}.). As a result, $\bar{\nu}_{\rm e}$ dominates over $\nu_{\rm e}$ in the forward direction alone and then the ELN crossing occurs.



At larger radii ($r \gtrsim 80$km), $N_{\nu_{\rm e}}$ and $N_{\bar{\nu}_{\rm e}}$ are gradually deviated from each other with increasing radius, since $\bar{\nu}_{\rm e}$ is more frequently absorbed or scattered by matter due to its higher average energy than that of $\nu_{\rm e}$. As a consequence, $\nu_{\rm e}$ dominates over $\bar{\nu}_{\rm e}$ in the forward direction again. We find, however, that $\bar{\nu}_{\rm e}$ dominates over $\nu_{\rm e}$ in the inward directions along the radial ray with $\theta = 45^{\circ}$ at $r \gtrsim 100$km (see green and purple lines in Fig.~\ref{graph_axisymmetric_crossingcheck_45deg_250ms}). It is due to the fact that the positron capture by neutrons is more frequent than the electron capture by protons, which can be seen in Fig.~\ref{graph_CompareECP_PCP_Back}, and the low-$Y_e$ environment is responsible (see the top panel in Fig.~\ref{graph_asymMatterprofile}). The negative $\mu_{\nu_{\rm e}}$ (see the solid red line in the bottom panel of Fig.~\ref{graph_asymMatterprofile}) is consistent with this interpretation. 




The low-$Y_e$ matter environment along a radial ray with $\theta = 45^{\circ}$ at $r \gtrsim 100$km is as a consequence of ejections of neutron-rich matter and higher $\bar{\nu}_{\rm e}$ emissions in the same hemisphere. We find that neutron-rich matter at $r \lesssim 100$km are dredged up by the neutrino-driven convections and, more interestingly, some of them are ejected, which is supposed to be associated with stronger shock expansion in the same hemisphere. Note that the higher $\bar{\nu}_{\rm e}$ emission than in the opposite hemisphere also provides a preferable condition to create low-$Y_e$ ejecta (see also \citet{2019MNRAS.488L.114F} for more details).



As shown in Fig.~\ref{graph_axisymmetric_crossingcheck_45deg_250ms}, the occurrence of ELN crossing is determined by the delicate balance of angular distributions between $\nu_{\rm e}$ and $\bar{\nu}_{\rm e}$. In such circumstances, asymmetric neutrino emissions affect the ELN angular distributions not only directly but also indirectly through the change of matter state. It should be noted, however, that they do not always play a positive role for the ELN crossing. Indeed, they tend to make the number densities of $\nu_{\rm e}$ and $\bar{\nu}_{\rm e}$ more different in the opposite hemisphere (the same hemisphere with a PNS proper motion), which is negative for the ELN crossing. Fig.~\ref{graph_axisymmetric_crossingcheck_135deg_250ms} vindicates this; $\nu_{\rm e}$ dominates over $\bar{\nu}_{\rm e}$ in all directions. This can be also seen in the first two panels from right on the top row in Fig.~\ref{SpatialDistri_Growth}, in which the stable region is widely spread in the same hemisphere.

We finally discuss the importance of non-axisymmetric properties of the angular distributions of neutrinos in momentum space. We note that there is no ELN crossings in the $\phi_{\nu}$-integrated-angular distribution at $r=75$km along a radial way with $\theta = 45^{\circ}$ (see blue lines in Fig.~\ref{graph_axisymmetric_crossingcheck_45deg_250ms}), although there exist unstable modes as shown in the right panels of Fig.~\ref{DR_GRdetail}. The unexpected result comes from the fact that the crossing is simply smeared out by the $\phi_{\nu}$ integration in Fig.~\ref{graph_axisymmetric_crossingcheck_45deg_250ms}, but it actually exists in the original distribution. In Fig.~\ref{3Dang} we display the ELN angular distributions with retaining the $\phi_{\nu}$ dependence for two spatial locations ($r=50, 75$km for the left- and right panels, respectively) at the same zenith angle $\theta = 45^{\circ}$ at $T_{\rm b} = 250$ms. The cyan and red colors imply the positive and negative ELN, respectively. As is clear in these plots, the red prolate-spheroids in both panels are tilted with respect to the radial direction, which is a noticeable sign of non-axisymmetry. In particular, it is inclined more strongly at $r=75$km than at $r=50$km, i.e., the non-axisymmetry is more remarkable in the former, and then the non-radial ELN crossing occurs as a consequence.


Such non-radial ELN crossings are not rare in fact. Fig.~\ref{graphCrosscheck} exhibits the type of ELN crossings by color on the 2D spatial map at $T_{\rm b} = 250$ms. The regions colored with red and blue denote the radial and non-radial ELN crossings, respectively. As shown in Fig.~\ref{graphCrosscheck}, non-radial ELN crossings occur in wide spatial ranges of both pre- and post-shock regions in this model. We speculate that the difference in lateral fluxes between $\nu_{\rm e}$ and $\bar{\nu}_{\rm e}$ is responsible for the non-radial ELN crossing. The lateral $\nu_{\rm e}$ flux tends to be negative sign, i.e., $\nu_{\rm e}$ advects from south to north, which is due to the fact that $\nu_{\rm e}$ emissions are stronger in the southern-hemisphere. On the other hand, $\bar{\nu}_{\rm e}$ has an opposite trend. This generates coherent lateral flux of ELN, and then induces the non-radial ELN crossings. It should be noted, however, that our Boltzmann solver suffers from the numerical diffusion in particular at larger radii as demonstrated in \citet{2017ApJ...847..133R}, which may artificially enhance the region of non-radial ELN crossing. For more quantitative discussions, we need to carry out higher-resolution simulations, which are beyond the scope of this paper, though. 

\section{Summary} \label{sec:summary}
In this paper, we conducted the linear stability analysis of the fast flavor collective neutrino oscillations based on a result of our latest axisymmetric CCSN model obtained with the full Boltzmann neutrino transport. In this model, we found remarkable asymmetric neutrino emissions associated with a non-spherical shock expansion and a PNS proper motion \citep{2019arXiv190704863N}. We reckoned that such coherent asymmetric neutrino emissions have an impact on the ELN crossing, affecting in turn the fast flavor conversion.

We found that there exist unstable modes in both pre- and post-shock regions in this model. In the former, the ELN crossing is not triggered by multi-dimensional effects but rather by coherent scatterings of neutrinos off heavy nuclei. Thanks to its higher average energy, $\bar{\nu}_{\rm e}$ experiences coherent scatterings of heavy nuclei more frequently than $\nu_{\rm e}$ and, as a result, $\bar{\nu}_{\rm e}$ is more abundant than $\nu_{\rm e}$ in the inward flight directions. Since $\nu_{\rm e}$ is dominant over $\bar{\nu}_{\rm e}$ in the forward directions, the ELN crossing occurs somewhere in between, which will trigger the fast flavor conversion. In the post-shock flows, on the other hand, we found that $\nu_{\rm e}$ dominates over $\bar{\nu}_{\rm e}$ in all directions until the initiation of non-spherical shock expansion accompanied by asymmetric neutrino emissions. Thereafter (from $T_{\rm b} \sim 200$ms on), however, we did find the ELN crossings and hence unstable modes in the hemisphere of higher $\bar{\nu}_{\rm e}$ emissions. The disparity in the number densities between $\nu_{\rm e}$ and $\bar{\nu}_{\rm e}$ is reduced by the anti-correlation of their number fluxes in the same hemisphere, and a preferable condition for the ELN crossing is produced. It should be also noted that asymmetric neutrino emissions were not observed in our previous CCSN model \citep{2018ApJ...854..136N}, which would be the main reason why \citet{2019PhRvD..99j3011D} found no ELN crossings in the post-shock region.


We then analyzed in detail the ELN crossings by closely inspecting the angular distributions of neutrinos in momentum space. It turns out that these ELN crossings have different origins. In the inner region (see, e.g., $r=50$km in Fig.~\ref{graph_axisymmetric_crossingcheck_45deg_250ms}), $\bar{\nu}_{\rm e}$ is dominant over $\nu_{\rm e}$ in the outward directions, whereas the trend is opposite in the inward directions. This happens because the enhanced emissions of $\bar{\nu}_{\rm e}$ in one hemisphere in this model makes the number densities of $\nu_{\rm e}$ and $\bar{\nu}_{\rm e}$ comparable to each other, while the angular distribution of $\bar{\nu}_{\rm e}$ is in general more forward-peaked owing to its earlier decoupling with matter. In the outer region (see, e.g., $r=100$ and $150$km in Fig.~\ref{graph_axisymmetric_crossingcheck_45deg_250ms}), on the other hand, the ELN crossing occurs in the opposite way: $\nu_{\rm e}$ is dominant over $\bar{\nu}_{\rm e}$ in the outward directions and vice versa in the inward directions. This occurs because the positron capture on free neutrons is more frequent than the electron capture by free protons in low-$Y_e$ environments, which results in the negative ELN in the inward direction and then inducing the ELN crossing. The low-$Y_e$ matter environment is as a consequence of ejections of neutron-rich matter and asymmetric neutrino emissions. The former would be associated with the non-spherical shock expansion and also being aided by the dredged-up by neutrino-driven convection at $r \lesssim 100$km. On the other hand, the latter in the hemisphere with higher $\bar{\nu}_{\rm e}$ emissions also provides a preferable condition to create low-$Y_e$ ejecta as discussed in \citet{2019MNRAS.488L.114F}. In contrast, the enhancement of higher $\nu_{\rm e}$ emissions in the opposite hemisphere suppresses the occurrence of ELN crossing, which implies that the asymmetric neutrino emissions can give rise to stabilize the fast flavor conversion there.


We also find that the non-radial ELN crossing occurs between the regions with no ELN crossing and the radial ELN crossing. The non-radial ELN crossing never happens in spherical symmetry, since the axisymmetry is imposed in the neutrino angular distribution in momentum space; hence, the non-radial ELN crossing is purely multi-dimensional effect. Indeed, the difference of lateral fluxes between $\nu_{\rm e}$ and $\bar{\nu}_{\rm e}$ is a primal cause of the non-radial ELN crossing. Whether it really occurs in CCSN core is a subtle problem, however, and the further studies with higher angular resolutions are needed to ascertain it. We will address the issue in the forthcoming paper.

As discussed in \citet{2019arXiv190704863N}, the asymmetric neutrino emissions observed in this model are correlated with the shock morphology and the NS kick. This implies that the occurrence of fast flavor conversion will be also correlated with them and may have strong impacts on observables such as nucleosynthetic yields and neutrino signals. As for the former, \citet{2019MNRAS.488L.114F} recently discussed the possible consequences in the explosive nucleosynthesis by asymmetric neutrino emissions. Our findings in this study shows a need for the further study of the impact of fast flavor conversions on their outcomes. As for the latter issue, the self-consistent CCSN simulations that take into account the fast flavor conversion somehow are required, which are one of the top priorities in our future project. We also note that the axisymmetric condition, which was imposed our CCSN model, may artificially enhance the asymmetry of neutrino emissions. This issue will be addressed once 3D CCSN simulations with full Boltzmann neutrino transport are available.

Finally, we make a few remarks. Our findings in this paper indicate that the enhancement of $\bar{\nu}_{\rm e}$ is a key to the occurrence of fast flavor conversion in the post-shock region, which was also pointed out by previous studies (see e.g., \citet{2018arXiv181206883A}). Importantly, such asymmetric neutrino emissions may be a common property in CCSNe; for instance, the lepton-emission self-sustained asymmetry, or LESA, appears in many 3D CCSN simulations regardless of numerical methods \citep{2014ApJ...792...96T,2018arXiv180910150G,2018arXiv181205738P,2018ApJ...865...81O,2019arXiv190608787V}. Moreover the fast flavor conversion is likely to occur commonly also in the pre-shock region \citep{2019arXiv190913131M} unless the number density of $\nu_{\rm e}$ is much larger than that of $\bar{\nu}_{\rm e}$. Other recent works (see e.g., \citet{2019arXiv190701002S}) also found the occurrence of collective neutrino oscillations in CCSNe. In order to treat all these phenomena more rigorously, the quantum kinetic treatment of neutrino transport should be explored further. The entire CCSN community will tackle its intricate problems including the aspect of technical issues more considerably in future, and will address them towards unveiling the explosion mechanism of CCSNe, although it may be a long way to go.








\acknowledgments 
 We acknowledge Sherwood Richers, Luke Johns, George Fuller and Adam Burrows for fruitful discussions. The numerical computations were performed on the supercomputers at K, at AICS, FX10 at Information Technology Center of Nagoya University. Large-scale storage of numerical data is supported by JLDG constructed over SINET4 of NII. H.N. was supported by Princeton University through DOE SciDAC4 Grant DE-SC0018297 (subaward 00009650). This work was also supported by Grant-in-Aid for the Scientific Research from the Ministry of Education, Culture, Sports, Science and Technology (MEXT), Japan (15K05093, 25870099, 26104006, 16H03986, 17H06357, 17H06365), HPCI Strategic Program of Japanese MEXT and K computer at the RIKEN (Project ID: hpci 160071, 160211, 170230, 170031, 170304, hp180179, hp180111, hp180239).
\bibliography{bibfile}

\begin{thebibliography}{}
\expandafter\ifx\csname natexlab\endcsname\relax\def\natexlab#1{#1}\fi
\providecommand{\url}[1]{\href{#1}{#1}}
\providecommand{\dodoi}[1]{doi:~\href{http://doi.org/#1}{\nolinkurl{#1}}}
\providecommand{\doeprint}[1]{\href{http://ascl.net/#1}{\nolinkurl{http://ascl.net/#1}}}
\providecommand{\doarXiv}[1]{\href{https://arxiv.org/abs/#1}{\nolinkurl{https://arxiv.org/abs/#1}}}

\bibitem[{{Abbar} \& {Duan}(2018)}]{2018PhRvD..98d3014A}
{Abbar}, S., \& {Duan}, H. 2018, \prd, 98, 043014,
  \dodoi{10.1103/PhysRevD.98.043014}

\bibitem[{{Abbar} {et~al.}(2018){Abbar}, {Duan}, {Sumiyoshi}, {Takiwaki}, \&
  {Volpe}}]{2018arXiv181206883A}
{Abbar}, S., {Duan}, H., {Sumiyoshi}, K., {Takiwaki}, T., \& {Volpe}, M.~C.
  2018, arXiv e-prints, arXiv:1812.06883.
\newblock \doarXiv{1812.06883}

\bibitem[{{Abbar} \& {Volpe}(2019{\natexlab{a}})}]{2019PhLB..790..545A}
{Abbar}, S., \& {Volpe}, M.~C. 2019{\natexlab{a}}, Physics Letters B, 790, 545,
  \dodoi{10.1016/j.physletb.2019.02.002}

\bibitem[{{Abbar} \& {Volpe}(2019{\natexlab{b}})}]{2018arXiv181104215A}
---. 2019{\natexlab{b}}, Physics Letters B, 790, 545,
  \dodoi{10.1016/j.physletb.2019.02.002}

\bibitem[{{Abe} {et~al.}(2011){Abe}, {Abe}, {Aihara}, {Fukuda}, {Hayato},
  {Huang}, {Ichikawa}, {Ikeda}, {Inoue}, {Ishino}, {Itow}, {Kajita}, {Kameda},
  {Kishimoto}, {Koga}, {Koshio}, {Lee}, {Minamino}, {Miura}, {Moriyama},
  {Nakahata}, {Nakamura}, {Nakaya}, {Nakayama}, {Nishijima}, {Nishimura},
  {Obayashi}, {Okumura}, {Sakuda}, {Sekiya}, {Shiozawa}, {Suzuki}, {Suzuki},
  {Takeda}, {Takeuchi}, {Tanaka}, {Tasaka}, {Tomura}, {Vagins}, {Wang}, \&
  {Yokoyama}}]{2011arXiv1109.3262A}
{Abe}, K., {Abe}, T., {Aihara}, H., {et~al.} 2011, arXiv e-prints,
  arXiv:1109.3262.
\newblock \doarXiv{1109.3262}

\bibitem[{{Airen} {et~al.}(2018){Airen}, {Capozzi}, {Chakraborty}, {Dasgupta},
  {Raffelt}, \& {Stirner}}]{2018JCAP...12..019A}
{Airen}, S., {Capozzi}, F., {Chakraborty}, S., {et~al.} 2018, Journal of
  Cosmology and Astro-Particle Physics, 2018, 019,
  \dodoi{10.1088/1475-7516/2018/12/019}

\bibitem[{{Bionta} {et~al.}(1987){Bionta}, {Blewitt}, {Bratton}, {Casper},
  {Ciocio}, {Claus}, {Cortez}, {Crouch}, {Dye}, {Errede}, {Foster}, {Gajewski},
  {Ganezer}, {Goldhaber}, {Haines}, {Jones}, {Kielczewska}, {Kropp}, {Learned},
  {Losecco}, {Matthews}, {Miller}, {Mudan}, {Park}, {Price}, {Reines},
  {Schultz}, {Seidel}, {Shumard}, {Sinclair}, {Sobel}, {Stone}, {Sulak},
  {Svoboda}, {Thornton}, {van der Velde}, \& {Wuest}}]{1987PhRvL..58.1494B}
{Bionta}, R.~M., {Blewitt}, G., {Bratton}, C.~B., {et~al.} 1987, \prl, 58,
  1494, \dodoi{10.1103/PhysRevLett.58.1494}

\bibitem[{{Bollig} {et~al.}(2017){Bollig}, {Janka}, {Lohs},
  {Mart{\'\i}nez-Pinedo}, {Horowitz}, \& {Melson}}]{2017PhRvL.119x2702B}
{Bollig}, R., {Janka}, H.~T., {Lohs}, A., {et~al.} 2017, \prl, 119, 242702,
  \dodoi{10.1103/PhysRevLett.119.242702}

\bibitem[{{Brandt} {et~al.}(2011){Brandt}, {Burrows}, {Ott}, \&
  {Livne}}]{2011ApJ...728....8B}
{Brandt}, T.~D., {Burrows}, A., {Ott}, C.~D., \& {Livne}, E. 2011, \apj, 728,
  8, \dodoi{10.1088/0004-637X/728/1/8}

\bibitem[{{Burrows} {et~al.}(2019){Burrows}, {Radice}, \&
  {Vartanyan}}]{2019MNRAS.tmp..538B}
{Burrows}, A., {Radice}, D., \& {Vartanyan}, D. 2019, \mnras, 538,
  \dodoi{10.1093/mnras/stz543}

\bibitem[{{Capozzi} {et~al.}(2017){Capozzi}, {Dasgupta}, {Lisi}, {Marrone}, \&
  {Mirizzi}}]{2017PhRvD..96d3016C}
{Capozzi}, F., {Dasgupta}, B., {Lisi}, E., {Marrone}, A., \& {Mirizzi}, A.
  2017, \prd, 96, 043016, \dodoi{10.1103/PhysRevD.96.043016}

\bibitem[{{Carlson} {et~al.}(1983){Carlson}, {Pandharipande}, \&
  {Wiringa}}]{1983NuPhA.401...59C}
{Carlson}, J., {Pandharipande}, V.~R., \& {Wiringa}, R.~B. 1983, Nuclear
  Physics A, 401, 59, \dodoi{10.1016/0375-9474(83)90336-6}

\bibitem[{{Chakraborty} {et~al.}(2016{\natexlab{a}}){Chakraborty}, {Hansen},
  {Izaguirre}, \& {Raffelt}}]{2016NuPhB.908..366C}
{Chakraborty}, S., {Hansen}, R., {Izaguirre}, I., \& {Raffelt}, G.
  2016{\natexlab{a}}, Nuclear Physics B, 908, 366,
  \dodoi{10.1016/j.nuclphysb.2016.02.012}

\bibitem[{{Chakraborty} {et~al.}(2016{\natexlab{b}}){Chakraborty}, {Hansen},
  {Izaguirre}, \& {Raffelt}}]{2016JCAP...03..042C}
{Chakraborty}, S., {Hansen}, R.~S., {Izaguirre}, I., \& {Raffelt}, G.~G.
  2016{\natexlab{b}}, Journal of Cosmology and Astro-Particle Physics, 2016,
  042, \dodoi{10.1088/1475-7516/2016/03/042}

\bibitem[{{Dasgupta} {et~al.}(2017){Dasgupta}, {Mirizzi}, \&
  {Sen}}]{2017JCAP...02..019D}
{Dasgupta}, B., {Mirizzi}, A., \& {Sen}, M. 2017, Journal of Cosmology and
  Astro-Particle Physics, 2017, 019, \dodoi{10.1088/1475-7516/2017/02/019}

\bibitem[{{Dasgupta} {et~al.}(2018){Dasgupta}, {Mirizzi}, \&
  {Sen}}]{2018PhRvD..98j3001D}
---. 2018, \prd, 98, 103001, \dodoi{10.1103/PhysRevD.98.103001}

\bibitem[{{Dasgupta} \& {Sen}(2018)}]{2018PhRvD..97b3017D}
{Dasgupta}, B., \& {Sen}, M. 2018, \prd, 97, 023017,
  \dodoi{10.1103/PhysRevD.97.023017}

\bibitem[{{Delfan Azari} {et~al.}(2019){Delfan Azari}, {Yamada}, {Morinaga},
  {Iwakami}, {Okawa}, {Nagakura}, \& {Sumiyoshi}}]{2019PhRvD..99j3011D}
{Delfan Azari}, M., {Yamada}, S., {Morinaga}, T., {et~al.} 2019, \prd, 99,
  103011, \dodoi{10.1103/PhysRevD.99.103011}

\bibitem[{{Dighe} \& {Smirnov}(2000)}]{2000PhRvD..62c3007D}
{Dighe}, A.~S., \& {Smirnov}, A.~Y. 2000, \prd, 62, 033007,
  \dodoi{10.1103/PhysRevD.62.033007}

\bibitem[{{Fujimoto} \& {Nagakura}(2019)}]{2019MNRAS.488L.114F}
{Fujimoto}, S.-i., \& {Nagakura}, H. 2019, \mnras, 488, L114,
  \dodoi{10.1093/mnrasl/slz111}

\bibitem[{{Furusawa} {et~al.}(2017{\natexlab{a}}){Furusawa}, {Nagakura},
  {Sumiyoshi}, {Kato}, \& {Yamada}}]{2017PhRvC..95b5809F}
{Furusawa}, S., {Nagakura}, H., {Sumiyoshi}, K., {Kato}, C., \& {Yamada}, S.
  2017{\natexlab{a}}, \prc, 95, 025809, \dodoi{10.1103/PhysRevC.95.025809}

\bibitem[{{Furusawa} {et~al.}(2017{\natexlab{b}}){Furusawa}, {Togashi},
  {Nagakura}, {Sumiyoshi}, {Yamada}, {Suzuki}, \&
  {Takano}}]{2017JPhG...44i4001F}
{Furusawa}, S., {Togashi}, H., {Nagakura}, H., {et~al.} 2017{\natexlab{b}},
  Journal of Physics G Nuclear Physics, 44, 094001,
  \dodoi{10.1088/1361-6471/aa7f35}

\bibitem[{{Glas} {et~al.}(2018){Glas}, {Janka}, {Melson}, {Stockinger}, \&
  {Just}}]{2018arXiv180910150G}
{Glas}, R., {Janka}, H.~T., {Melson}, T., {Stockinger}, G., \& {Just}, O. 2018,
  arXiv e-prints, arXiv:1809.10150.
\newblock \doarXiv{1809.10150}

\bibitem[{{Harada} {et~al.}(2019){Harada}, {Nagakura}, {Iwakami}, {Okawa},
  {Furusawa}, {Matsufuru}, {Sumiyoshi}, \& {Yamada}}]{2019ApJ...872..181H}
{Harada}, A., {Nagakura}, H., {Iwakami}, W., {et~al.} 2019, \apj, 872, 181,
  \dodoi{10.3847/1538-4357/ab0203}

\bibitem[{{Hirata} {et~al.}(1987){Hirata}, {Kajita}, {Koshiba}, {Nakahata},
  {Oyama}, {Sato}, {Suzuki}, {Takita}, {Totsuka}, {Kifune}, {Suda},
  {Takahashi}, {Tanimori}, {Miyano}, {Yamada}, {Beier}, {Feldscher}, {Kim},
  {Mann}, {Newcomer}, {van}, {Zhang}, \& {Cortez}}]{1987PhRvL..58.1490H}
{Hirata}, K., {Kajita}, T., {Koshiba}, M., {et~al.} 1987, \prl, 58, 1490,
  \dodoi{10.1103/PhysRevLett.58.1490}

\bibitem[{{Horiuchi} \& {Kneller}(2018)}]{2018JPhG...45d3002H}
{Horiuchi}, S., \& {Kneller}, J.~P. 2018, Journal of Physics G Nuclear Physics,
  45, 043002, \dodoi{10.1088/1361-6471/aaa90a}

\bibitem[{{Hyper-Kamiokande Proto-Collaboration}
  {et~al.}(2018){Hyper-Kamiokande Proto-Collaboration}, {:}, {Abe}, {Abe},
  {Aihara}, {Aimi}, {Akutsu}, {Andreopoulos}, {Anghel}, {Anthony}, {Antonova},
  {Ashida}, {Aushev}, {Barbi}, {Barker}, {Barr}, {Beltrame}, {Berardi},
  {Bergevin}, {Berkman}, {Berns}, {Berry}, {Bhadra}, {Bravo-Bergu{\~n}o},
  {Blaszczyk}, {Blondel}, {Bolognesi}, {Boyd}, {Bravar}, {Bronner}, {Buizza
  Avanzini}, {Cafagna}, {Cole}, {Calland}, {Cao}, {Cartwright}, {Catanesi},
  {Checchia}, {Chen-Wishart}, {Choi}, {Choi}, {Coleman}, {Collazuol}, {Cowan},
  {Cremonesi}, {Dealtry}, {De Rosa}, {Densham}, {Dewhurst}, {Drakopoulou}, {Di
  Lodovico}, {Drapier}, {Dumarchez}, {Dunne}, {Dziewiecki}, {Emery}, {Esmaili},
  {Evangelisti}, {Fernandez-Martinez}, {Feusels}, {Finch}, {Fiorentini},
  {Fiorillo}, {Fitton}, {Frankiewicz}, {Friend}, {Fujii}, {Fukuda}, {Fukuda},
  {Ganezer}, {Giganti}, {Gonin}, {Grant}, {Gumplinger}, {Hadley}, {Hartfiel},
  {Hartz}, {Hayato}, {Hayrapetyan}, {Hill}, {Hirota}, {Horiuchi}, {Ichikawa},
  {Iijima}, {Ikeda}, {Imber}, {Inoue}, {Insler}, {Intonti}, {Ioannisian},
  {Ishida}, {Ishino}, {Ishitsuka}, {Itow}, {Iwamoto}, {Izmaylov}, {Jamieson},
  {Jang}, {Jang}, {Jeon}, {Jiang}, {Jonsson}, {Joo}, {Kaboth}, {Kachulis},
  {Kajita}, {Kameda}, {Kataoka}, {Katori}, {Kayrapetyan}, {Kearns},
  {Khabibullin}, {Khotjantsev}, {Kim}, {Kim}, {Kim}, {Kim}, {King},
  {Kishimoto}, {Kobayashi}, {Koga}, {Konaka}, {Kormos}, {Koshio}, {Korzenev},
  {Kowalik}, {Kropp}, {Kudenko}, {Kurjata}, {Kutter}, {Kuze}, {Labarga},
  {Lagoda}, {Lasorak}, {Laveder}, {Lawe}, {Learned}, {Lim}, {Lindner},
  {Litchfield}, {Longhin}, {Loverre}, {Lou}, {Ludovici}, {Ma}, {Magaletti},
  {Mahn}, {Malek}, {Maret}, {Mariani}, {Martens}, {Marti}, {Martin}, {Marzec},
  {Matsuno}, {Mazzucato}, {McCarthy}, {McCauley}, {McFarland }, {McGrew},
  {Mefodiev}, {Mermod}, {Metelko}, {Mezzetto}, {Migenda}, {Mijakowski},
  {Minakata}, {Minamino}, {Mine}, {Mineev}, {Mitra}, {Miura}, {Mochizuki},
  {Monroe}, {Moon}, {Moriyama}, {Mueller}, {Muheim}, {Murase}, {Muto},
  {Nakahata}, {Nakajima}, {Nakamura}, {Nakaya}, {Nakayama}, {Nantais},
  {Needham}, {Nicholls}, {Nishimura}, {Noah}, {Nova}, {Nowak}, {Nunokawa},
  {Obayashi}, {O'Keeffe}, {Okajima}, {Okumura}, {Onishchuk}, {O'Sullivan},
  {O'Sullivan}, {Ovsiannikova}, {Owen}, {Oyama}, {Pac}, {Palladino},
  {Palomino}, {Paolone}, {Parker}, {Parsa}, {Payne}, {Perkin}, {Pidcott},
  {Pinzon Guerra}, {Playfer}, {Popov}, {Posiadala-Zezula}, {Poutissou},
  {Pritchard}, {Prouse}, {Pronost}, {Przewlocki}, {Quilain}, {Radicioni},
  {Ratoff}, {Retiere}, {Riccio}, {Richards}, {Rondio}, {Rose}, {Rott},
  {Rountree}, {Ruggeri}, {Rychter}, {Sacco}, {Sakuda}, {Sanchez},
  {Scantamburlo}, {Scott}, {Sedgwick}, {Seiya}, {Sekiguchi}, {Sekiya}, {Seo},
  {Sgalaberna}, {Shah}, {Shaikhiev}, {Shimizu}, {Shiozawa}, {Shitov}, {Short},
  {Simpson}, {Sinnis}, {Smy}, {Snow}, {Sobczyk}, {Sobel}, {Sonoda}, {Spina},
  {Stewart}, {Stone}, {Suda}, {Suwa}, {Suzuki}, {Suzuki}, {Svoboda}, {Taani},
  {Tacik}, {Takeda}, {Takenaka}, {Taketa}, {Takeuchi}, {Takhistov}, {Tanaka},
  {Tanaka}, {Tanaka}, {Terri}, {Thiesse}, {Thompson}, {Thorpe}, {Tobayama},
  {Touramanis}, {Towstego}, {Tsukamoto}, {Tsui}, {Tzanov}, {Uchida}, {Vagins},
  {Vasseur}, {Vilela}, {Vogelaar}, {Walding}, {Walker}, {Ward}, {Wark},
  {Wascko}, {Weber}, {Wendell}, {Wilkes}, {Wilking}, {Wilson}, {Xin},
  {Yamamoto}, {Yanagisawa}, {Yano}, {Yen}, {Yershov}, {Yeum}, {Yokoyama},
  {Yoshida}, {Yu}, {Yu}, {Zalipska}, {Zaremba}, {Ziembicki}, {Zito}, \&
  {Zsoldos}}]{2018arXiv180504163H}
{Hyper-Kamiokande Proto-Collaboration}, {:}, {Abe}, K., {et~al.} 2018, arXiv
  e-prints, arXiv:1805.04163.
\newblock \doarXiv{1805.04163}

\bibitem[{{Izaguirre} {et~al.}(2017){Izaguirre}, {Raffelt}, \&
  {Tamborra}}]{2017PhRvL.118b1101I}
{Izaguirre}, I., {Raffelt}, G., \& {Tamborra}, I. 2017, \prl, 118, 021101,
  \dodoi{10.1103/PhysRevLett.118.021101}

\bibitem[{{Kotake} {et~al.}(2012){Kotake}, {Sumiyoshi}, {Yamada}, {Takiwaki},
  {Kuroda}, {Suwa}, \& {Nagakura}}]{2012PTEP.2012aA301K}
{Kotake}, K., {Sumiyoshi}, K., {Yamada}, S., {et~al.} 2012, Progress of
  Theoretical and Experimental Physics, 2012, 01A301,
  \dodoi{10.1093/ptep/pts009}

\bibitem[{{Kuroda} {et~al.}(2016){Kuroda}, {Takiwaki}, \&
  {Kotake}}]{2016ApJS..222...20K}
{Kuroda}, T., {Takiwaki}, T., \& {Kotake}, K. 2016, \apjs, 222, 20,
  \dodoi{10.3847/0067-0049/222/2/20}

\bibitem[{{Lentz} {et~al.}(2015){Lentz}, {Bruenn}, {Hix}, {Mezzacappa},
  {Messer}, {Endeve}, {Blondin}, {Harris}, {Marronetti}, \&
  {Yakunin}}]{2015ApJ...807L..31L}
{Lentz}, E.~J., {Bruenn}, S.~W., {Hix}, W.~R., {et~al.} 2015, \apjl, 807, L31,
  \dodoi{10.1088/2041-8205/807/2/L31}

\bibitem[{{Melson} \& {Janka}(2019)}]{2019arXiv190401699M}
{Melson}, T., \& {Janka}, H.~T. 2019, arXiv e-prints, arXiv:1904.01699.
\newblock \doarXiv{1904.01699}

\bibitem[{{Melson} {et~al.}(2015){Melson}, {Janka}, {Bollig}, {Hanke}, {Marek},
  \& {M{\"u}ller}}]{2015ApJ...808L..42M}
{Melson}, T., {Janka}, H.-T., {Bollig}, R., {et~al.} 2015, \apj, 808, L42,
  \dodoi{10.1088/2041-8205/808/2/L42}

\bibitem[{{Mirizzi} {et~al.}(2016){Mirizzi}, {Tamborra}, {Janka}, {Saviano},
  {Scholberg}, {Bollig}, {H{\"u}depohl}, \&
  {Chakraborty}}]{2016NCimR..39....1M}
{Mirizzi}, A., {Tamborra}, I., {Janka}, H.~T., {et~al.} 2016, Nuovo Cimento
  Rivista Serie, 39, 1, \dodoi{10.1393/ncr/i2016-10120-8}

\bibitem[{{Morinaga} {et~al.}(2019){Morinaga}, {Nagakura}, {Kato}, \&
  {Yamada}}]{2019arXiv190913131M}
{Morinaga}, T., {Nagakura}, H., {Kato}, C., \& {Yamada}, S. 2019, arXiv
  e-prints, arXiv:1909.13131.
\newblock \doarXiv{1909.13131}

\bibitem[{{Morinaga} \& {Yamada}(2018)}]{2018PhRvD..97b3024M}
{Morinaga}, T., \& {Yamada}, S. 2018, \prd, 97, 023024,
  \dodoi{10.1103/PhysRevD.97.023024}

\bibitem[{{M{\"u}ller} {et~al.}(2017){M{\"u}ller}, {Melson}, {Heger}, \&
  {Janka}}]{2017MNRAS.472..491M}
{M{\"u}ller}, B., {Melson}, T., {Heger}, A., \& {Janka}, H.-T. 2017, \mnras,
  472, 491, \dodoi{10.1093/mnras/stx1962}

\bibitem[{{M{\"u}ller} {et~al.}(2018){M{\"u}ller}, {Tauris}, {Heger},
  {Banerjee}, {Qian}, {Powell}, {Chan}, {Gay}, \&
  {Langer}}]{2018arXiv181105483M}
{M{\"u}ller}, B., {Tauris}, T.~M., {Heger}, A., {et~al.} 2018, ArXiv e-prints.
\newblock \doarXiv{1811.05483}

\bibitem[{{Nagakura} {et~al.}(2019{\natexlab{a}}){Nagakura}, {Burrows},
  {Radice}, \& {Vartanyan}}]{2019arXiv190503786N}
{Nagakura}, H., {Burrows}, A., {Radice}, D., \& {Vartanyan}, D.
  2019{\natexlab{a}}, arXiv e-prints, arXiv:1905.03786.
\newblock \doarXiv{1905.03786}

\bibitem[{{Nagakura} {et~al.}(2019{\natexlab{b}}){Nagakura}, {Furusawa},
  {Togashi}, {Richers}, {Sumiyoshi}, \& {Yamada}}]{2019ApJS..240...38N}
{Nagakura}, H., {Furusawa}, S., {Togashi}, H., {et~al.} 2019{\natexlab{b}},
  \apjs, 240, 38, \dodoi{10.3847/1538-4365/aafac9}

\bibitem[{{Nagakura} {et~al.}(2017){Nagakura}, {Iwakami}, {Furusawa},
  {Sumiyoshi}, {Yamada}, {Matsufuru}, \& {Imakura}}]{2017ApJS..229...42N}
{Nagakura}, H., {Iwakami}, W., {Furusawa}, S., {et~al.} 2017, \apjs, 229, 42,
  \dodoi{10.3847/1538-4365/aa69ea}

\bibitem[{{Nagakura} {et~al.}(2014){Nagakura}, {Sumiyoshi}, \&
  {Yamada}}]{2014ApJS..214...16N}
{Nagakura}, H., {Sumiyoshi}, K., \& {Yamada}, S. 2014, \apjs, 214, 16,
  \dodoi{10.1088/0067-0049/214/2/16}

\bibitem[{{Nagakura} {et~al.}(2019{\natexlab{c}}){Nagakura}, {Sumiyoshi}, \&
  {Yamada}}]{2019arXiv190704863N}
---. 2019{\natexlab{c}}, \apjl, 880, L28, \dodoi{10.3847/2041-8213/ab30ca}

\bibitem[{{Nagakura} {et~al.}(2019{\natexlab{d}}){Nagakura}, {Sumiyoshi}, \&
  {Yamada}}]{2019arXiv190610143N}
---. 2019{\natexlab{d}}, \apj, 878, 160, \dodoi{10.3847/1538-4357/ab2189}

\bibitem[{{Nagakura} {et~al.}(2018){Nagakura}, {Iwakami}, {Furusawa}, {Okawa},
  {Harada}, {Sumiyoshi}, {Yamada}, {Matsufuru}, \&
  {Imakura}}]{2018ApJ...854..136N}
{Nagakura}, H., {Iwakami}, W., {Furusawa}, S., {et~al.} 2018, \apj, 854, 136,
  \dodoi{10.3847/1538-4357/aaac29}

\bibitem[{{Nakamura} {et~al.}(2016){Nakamura}, {Horiuchi}, {Tanaka}, {Hayama},
  {Takiwaki}, \& {Kotake}}]{2016MNRAS.461.3296N}
{Nakamura}, K., {Horiuchi}, S., {Tanaka}, M., {et~al.} 2016, \mnras, 461, 3296,
  \dodoi{10.1093/mnras/stw1453}

\bibitem[{{Nakamura} {et~al.}(2019){Nakamura}, {Takiwaki}, \&
  {Kotake}}]{2019arXiv190408088N}
{Nakamura}, K., {Takiwaki}, T., \& {Kotake}, K. 2019, arXiv e-prints,
  arXiv:1904.08088.
\newblock \doarXiv{1904.08088}

\bibitem[{{O'Connor} \& {Couch}(2018)}]{2018ApJ...865...81O}
{O'Connor}, E.~P., \& {Couch}, S.~M. 2018, \apj, 865, 81,
  \dodoi{10.3847/1538-4357/aadcf7}

\bibitem[{{Ott} {et~al.}(2008){Ott}, {Burrows}, {Dessart}, \&
  {Livne}}]{2008ApJ...685.1069O}
{Ott}, C.~D., {Burrows}, A., {Dessart}, L., \& {Livne}, E. 2008, \apj, 685,
  1069, \dodoi{10.1086/591440}

\bibitem[{{Ott} {et~al.}(2018){Ott}, {Roberts}, {da Silva Schneider}, {Fedrow},
  {Haas}, \& {Schnetter}}]{2018ApJ...855L...3O}
{Ott}, C.~D., {Roberts}, L.~F., {da Silva Schneider}, A., {et~al.} 2018, \apjl,
  855, L3, \dodoi{10.3847/2041-8213/aaa967}

\bibitem[{{Powell} \& {M{\"u}ller}(2018)}]{2018arXiv181205738P}
{Powell}, J., \& {M{\"u}ller}, B. 2018, arXiv e-prints, arXiv:1812.05738.
\newblock \doarXiv{1812.05738}

\bibitem[{{Pudliner} {et~al.}(1995){Pudliner}, {Pandharipande}, {Carlson}, \&
  {Wiringa}}]{1995PhRvL..74.4396P}
{Pudliner}, B.~S., {Pandharipande}, V.~R., {Carlson}, J., \& {Wiringa}, R.~B.
  1995, Physical Review Letters, 74, 4396, \dodoi{10.1103/PhysRevLett.74.4396}

\bibitem[{{Raffelt}(2011)}]{2011NuPhS.221..218R}
{Raffelt}, G.~G. 2011, Nuclear Physics B Proceedings Supplements, 221, 218,
  \dodoi{10.1016/j.nuclphysbps.2011.09.006}

\bibitem[{{Richers} {et~al.}(2017){Richers}, {Nagakura}, {Ott}, {Dolence},
  {Sumiyoshi}, \& {Yamada}}]{2017ApJ...847..133R}
{Richers}, S., {Nagakura}, H., {Ott}, C.~D., {et~al.} 2017, \apj, 847, 133,
  \dodoi{10.3847/1538-4357/aa8bb2}

\bibitem[{{Richers} {et~al.}(2019){Richers}, {McLaughlin}, {Kneller}, \&
  {Vlasenko}}]{2019arXiv190300022R}
{Richers}, S.~A., {McLaughlin}, G.~C., {Kneller}, J.~P., \& {Vlasenko}, A.
  2019, arXiv e-prints, arXiv:1903.00022.
\newblock \doarXiv{1903.00022}

\bibitem[{{Roberts} {et~al.}(2016){Roberts}, {Ott}, {Haas}, {O'Connor},
  {Diener}, \& {Schnetter}}]{2016ApJ...831...98R}
{Roberts}, L.~F., {Ott}, C.~D., {Haas}, R., {et~al.} 2016, \apj, 831, 98,
  \dodoi{10.3847/0004-637X/831/1/98}

\bibitem[{{Sarikas} {et~al.}(2012){Sarikas}, {de Sousa Seixas}, \&
  {Raffelt}}]{2012PhRvD..86l5020S}
{Sarikas}, S., {de Sousa Seixas}, D., \& {Raffelt}, G. 2012, \prd, 86, 125020,
  \dodoi{10.1103/PhysRevD.86.125020}

\bibitem[{{Sasaki} {et~al.}(2019){Sasaki}, {Takiwaki}, {Kawagoe}, {Horiuchi},
  \& {Ishidoshiro}}]{2019arXiv190701002S}
{Sasaki}, H., {Takiwaki}, T., {Kawagoe}, S., {Horiuchi}, S., \& {Ishidoshiro},
  K. 2019, arXiv e-prints, arXiv:1907.01002.
\newblock \doarXiv{1907.01002}

\bibitem[{{Sawyer}(2005)}]{2005PhRvD..72d5003S}
{Sawyer}, R.~F. 2005, \prd, 72, 045003, \dodoi{10.1103/PhysRevD.72.045003}

\bibitem[{{Sawyer}(2016)}]{2016PhRvL.116h1101S}
---. 2016, \prl, 116, 081101, \dodoi{10.1103/PhysRevLett.116.081101}

\bibitem[{{Seadrow} {et~al.}(2018){Seadrow}, {Burrows}, {Vartanyan}, {Radice},
  \& {Skinner}}]{2018MNRAS.480.4710S}
{Seadrow}, S., {Burrows}, A., {Vartanyan}, D., {Radice}, D., \& {Skinner},
  M.~A. 2018, \mnras, 480, 4710, \dodoi{10.1093/mnras/sty2164}

\bibitem[{{Shalgar} \& {Tamborra}(2019)}]{2019arXiv190407236S}
{Shalgar}, S., \& {Tamborra}, I. 2019, arXiv e-prints, arXiv:1904.07236.
\newblock \doarXiv{1904.07236}

\bibitem[{{Sumiyoshi} \& {Yamada}(2012)}]{2012ApJS..199...17S}
{Sumiyoshi}, K., \& {Yamada}, S. 2012, \apjs, 199, 17,
  \dodoi{10.1088/0067-0049/199/1/17}

\bibitem[{{Summa} {et~al.}(2018){Summa}, {Janka}, {Melson}, \&
  {Marek}}]{2018ApJ...852...28S}
{Summa}, A., {Janka}, H.-T., {Melson}, T., \& {Marek}, A. 2018, \apj, 852, 28,
  \dodoi{10.3847/1538-4357/aa9ce8}

\bibitem[{{Suwa} {et~al.}(2019){Suwa}, {Sumiyoshi}, {Nakazato}, {Takahira},
  {Koshio}, {Mori}, \& {Wendell}}]{2019arXiv190409996S}
{Suwa}, Y., {Sumiyoshi}, K., {Nakazato}, K., {et~al.} 2019, arXiv e-prints,
  arXiv:1904.09996.
\newblock \doarXiv{1904.09996}

\bibitem[{{Takiwaki} {et~al.}(2012){Takiwaki}, {Kotake}, \&
  {Suwa}}]{2012ApJ...749...98T}
{Takiwaki}, T., {Kotake}, K., \& {Suwa}, Y. 2012, \apj, 749, 98,
  \dodoi{10.1088/0004-637X/749/2/98}

\bibitem[{{Takiwaki} {et~al.}(2014){Takiwaki}, {Kotake}, \&
  {Suwa}}]{2014ApJ...786...83T}
---. 2014, \apj, 786, 83, \dodoi{10.1088/0004-637X/786/2/83}

\bibitem[{{Takiwaki} {et~al.}(2016){Takiwaki}, {Kotake}, \&
  {Suwa}}]{2016MNRAS.461L.112T}
---. 2016, \mnras, 461, L112, \dodoi{10.1093/mnrasl/slw105}

\bibitem[{{Tamborra} {et~al.}(2014){Tamborra}, {Hanke}, {Janka}, {M{\"u}ller},
  {Raffelt}, \& {Marek}}]{2014ApJ...792...96T}
{Tamborra}, I., {Hanke}, F., {Janka}, H.-T., {et~al.} 2014, \apj, 792, 96,
  \dodoi{10.1088/0004-637X/792/2/96}

\bibitem[{{Tamborra} {et~al.}(2017){Tamborra}, {H{\"u}depohl}, {Raffelt}, \&
  {Janka}}]{2017ApJ...839..132T}
{Tamborra}, I., {H{\"u}depohl}, L., {Raffelt}, G.~G., \& {Janka}, H.-T. 2017,
  \apj, 839, 132, \dodoi{10.3847/1538-4357/aa6a18}

\bibitem[{{Tian} {et~al.}(2017){Tian}, {Patwardhan}, \&
  {Fuller}}]{2017PhRvD..95f3004T}
{Tian}, J.~Y., {Patwardhan}, A.~V., \& {Fuller}, G.~M. 2017, \prd, 95, 063004,
  \dodoi{10.1103/PhysRevD.95.063004}

\bibitem[{{Togashi} {et~al.}(2017){Togashi}, {Nakazato}, {Takehara},
  {Yamamuro}, {Suzuki}, \& {Takano}}]{2017NuPhA.961...78T}
{Togashi}, H., {Nakazato}, K., {Takehara}, Y., {et~al.} 2017, Nuclear Physics
  A, 961, 78, \dodoi{10.1016/j.nuclphysa.2017.02.010}

\bibitem[{{Togashi} \& {Takano}(2013)}]{2013NuPhA.902...53T}
{Togashi}, H., \& {Takano}, M. 2013, Nuclear Physics A, 902, 53,
  \dodoi{10.1016/j.nuclphysa.2013.02.014}

\bibitem[{{Vartanyan} {et~al.}(2019{\natexlab{a}}){Vartanyan}, {Burrows}, \&
  {Radice}}]{2019arXiv190608787V}
{Vartanyan}, D., {Burrows}, A., \& {Radice}, D. 2019{\natexlab{a}}, arXiv
  e-prints, arXiv:1906.08787.
\newblock \doarXiv{1906.08787}

\bibitem[{{Vartanyan} {et~al.}(2019{\natexlab{b}}){Vartanyan}, {Burrows},
  {Radice}, {Skinner}, \& {Dolence}}]{2019MNRAS.482..351V}
{Vartanyan}, D., {Burrows}, A., {Radice}, D., {Skinner}, M.~A., \& {Dolence},
  J. 2019{\natexlab{b}}, \mnras, 482, 351, \dodoi{10.1093/mnras/sty2585}

\bibitem[{{Wiringa} {et~al.}(1995){Wiringa}, {Stoks}, \&
  {Schiavilla}}]{1995PhRvC..51...38W}
{Wiringa}, R.~B., {Stoks}, V.~G.~J., \& {Schiavilla}, R. 1995, \prc, 51, 38,
  \dodoi{10.1103/PhysRevC.51.38}

\bibitem[{{Wolfenstein}(1979)}]{1979PhRvD..20.2634W}
{Wolfenstein}, L. 1979, \prd, 20, 2634, \dodoi{10.1103/PhysRevD.20.2634}

\bibitem[{{Woosley} {et~al.}(2002){Woosley}, {Heger}, \&
  {Weaver}}]{2002RvMP...74.1015W}
{Woosley}, S.~E., {Heger}, A., \& {Weaver}, T.~A. 2002, Reviews of Modern
  Physics, 74, 1015, \dodoi{10.1103/RevModPhys.74.1015}

\bibitem[{{Yi} {et~al.}(2019){Yi}, {Ma}, {Martin}, \&
  {Duan}}]{2019PhRvD..99f3005Y}
{Yi}, C., {Ma}, L., {Martin}, J.~D., \& {Duan}, H. 2019, \prd, 99, 063005,
  \dodoi{10.1103/PhysRevD.99.063005}

\end{thebibliography}

\end{document}